\documentclass[useAMS,usenatbib]{mn2e}
\usepackage{psfig}
\usepackage{amssymb}
\usepackage{color}
\usepackage{enumerate}
\usepackage{rotating}
\usepackage{amsmath}
\usepackage{mathrsfs}
\usepackage{ragged2e}

\title[The High-Mass RS from BOSS]
{The High-Mass End of the Red Sequence at $z\sim0.55$ from SDSS-III/BOSS: completeness, bimodality and luminosity function \vspace{-0.5cm}}

\author[Montero-Dorta et al.]{
\parbox[t]{\textwidth}{
Antonio D. Montero-Dorta$^{1}$\thanks{E-mail: amontero@astro.utah.edu}, Adam S. Bolton$^{1}$, Joel R. Brownstein$^{1}$, Molly Swanson$^5$, 
Kyle Dawson$^{1}$,  Francisco Prada$^{2,3,4}$, Daniel Eisenstein$^5$, Claudia Maraston$^{6}$, Daniel Thomas$^{6}$,
Johan Comparat$^{3,4}$, Chia-Hsun Chuang$^{3,4}$, Cameron K. McBride$^5$, Ginevra Favole$^{3,4}$, Hong Guo$^{9}$, Sergio Rodr\'iguez-Torres$^{3,4}$, Donald P. Schneider$^{7,8}$}
\vspace*{6pt} \\ 
$^1$ Department of Physics and Astronomy, The University of Utah, 115
South 1400 East, Salt Lake City, UT 84112, USA \\
$^2$ Instituto de Astrof\'isica de Andaluc\'ia (CSIC), Granada, E-18008, Spain \\
$^3$ Campus of International Excellence UAM+CSIC, Cantoblanco, E-28049 Madrid, Spain\\
$^4$ Instituto de F\'{\i}sica Te\'orica, (UAM/CSIC), Universidad Aut\'onoma de Madrid, Cantoblanco, E-28049 Madrid, Spain\\
$^5$ Harvard-Smithsonian Center for Astrophysics, 60 Garden St., Cambridge, MA 02138\\
$^6$ Institute of Cosmology \& Gravitation, University of Portsmouth, Dennis Sciama Building, Portsmouth PO1 3FX, UK \\
$^7$ Department of Astronomy and Astrophysics, The Pennsylvania State University, University Park, PA 16802 \\
$^8$ Institute for Gravitation and the Cosmos, The Pennsylvania State University, University Park, PA 16802 \\
$^9$ Key Laboratory for Research in Galaxies and Cosmology, Shanghai Astronomical Observatory, Shanghai 200030, China
\vspace{-0.4cm} 
}

\date{Accepted ---. Received ---;in original form --- \vspace{-0.3cm}}

\def\simlt{\lower.5ex\hbox{$\; \buildrel < \over \sim \;$}}
\def\simgt{\lower.5ex\hbox{$\; \buildrel > \over \sim \;$}}
\usepackage{graphicx}
\usepackage{rotating}

\definecolor{red}{rgb}{1,0,0}

\begin{document}

\bibliographystyle{mnras}

\maketitle

\begin{abstract}

We have developed an analytical method based on forward-modeling techniques to characterize the high-mass end  
of the red sequence (RS) galaxy population at redshift $z\sim0.55$, from the DR10 BOSS CMASS spectroscopic sample, 
which comprises $\sim600,000$ galaxies. The method, which follows an unbinned maximum 
likelihood approach, allows the deconvolution of the intrinsic CMASS colour-colour-magnitude distributions 
from photometric errors and selection effects. This procedure requires modeling the covariance matrix for the i-band magnitude, 
g-r colour and r-i colour using Stripe 82 multi-epoch data. Our results indicate that the error-deconvolved intrinsic RS distribution is consistent, 
within the photometric uncertainties, with a single point ($<0.05~{\rm{mag}}$) in the colour-colour plane at fixed magnitude, for a narrow redshift slice.
We have computed the high-mass end ($^{0.55}M_i \lesssim -22$) of the $^{0.55}i$-band RS Luminosity Function (RS LF) in several redshift slices 
within the redshift range $0.52<z<0.63$. In this narrow redshift range, the evolution of the RS LF is consistent, within the uncertainties in the 
modeling, with a passively-evolving model with $\Phi_* = (7.248 \pm 0.204) \times10^{-4}$ Mpc$^{-3}$ mag$^{-1}$, 
fading at a rate of $1.5\pm0.4$ mag per unit redshift. We report RS completeness as a function of magnitude and redshift in the CMASS sample,
which will facilitate a variety of galaxy-evolution and clustering studies using BOSS. Our forward-modeling 
method lays the foundations for future studies using other dark-energy surveys like eBOSS or DESI, which are 
affected by the same type of photometric blurring/selection effects.

\end{abstract}


\section{Introduction}
\label{sec:intro}

At low redshift, large spectroscopic surveys like the Sloan Digital Sky Survey (SDSS, \citealt{York2000}) or 
the Two Degree Field Galaxy Redshift Survey (2dFGRS, \citealt{Colless2001}) have allowed a detailed characterization 
of the main statistical properties of the galaxy population, serving as a solid benchmark for galaxy-evolution and 
large-scale-structure (LSS)/clustering studies. At higher redshift, however, establishing these mean statistical properties has proven significantly 
challenging, as galaxy samples are strongly affected by low-number statistics and selection effects. 

In recent years, a new type of survey has emerged, increasing the number of galaxies accessible 
for population studies at higher redshifts. The so-called dark-energy (DE) surveys 
collect data for millions of galaxies with the specific goal of understanding the nature of the DE
 that drives the present-day accelerated expansion of the Universe. The largest of these
DE experiments to date is the Baryon Oscillation Spectroscopic Survey (BOSS, \citealt{Schlegel2009a}, \citealt{Dawson2013}) of 
the SDSS-III \citep{Eisenstein2011}. BOSS uses a sample of over $1.5$ million luminous red galaxies 
(LRGs) to measure the Baryon Acoustic Oscillation (BAO, \citealt{Eisenstein2001}) at $z\simeq0.55$. 
The BAO is an overdensity of baryonic matter that can be used as a {\it{standard ruler}} for measuring 
the acceleration of the Universe and, consequently, exploring the nature of dark 
energy (e.g. \citealt{Eisenstein2001,Eisenstein2005, Drinkwater2010, Blake2011}).

The unprecedented statistics offered by BOSS, and other future DE surveys, 
have also great potential to clarify the massive galaxy evolution picture at $z < 1$. 
Realizing such potential presents, however, significant challenges. While the current state-of-the-art 
of {\it{precision cosmology}} dictates the use of huge volumes and large densities, 
in terms of redshift estimation, low signal-to-noise (S/N) spectra are generally sufficient. In addition, the effect of 
photometric uncertainties becomes progressively more severe, as we explore higher and higher redshifts ranges.
This paper is aimed at developing the adequate statistical tools to maximize the amount of 
galaxy-population information that we can extract from BOSS, in order to 
shed light into the massive end ($M_* \gtrsim 10^{11}~M_\odot$) of the red sequence (RS) at $z\sim0.55$, 
and to increase the accuracy of cosmological analyses that use these galaxies as tracers of the LSS.

The evolution of $L_*$ galaxies from $z\sim1$ have been thoroughly studied using 
a variety of galaxy samples (often combining different redshift surveys). It appears well stablished that the number density for blue $L_*$ 
galaxies remains fairly constant at $z<1$, whereas the number density of red $L_*$ galaxies experiences a significant increase over
the same period of time ($\Phi_*$ increases at least by a factor 2, according to \citealt{Faber2007}). The very massive end of the 
red sequence population, the main focus of this work, has been hard to probe, however, mostly due to footprint limitations. There is, nevertheless, clear indications
of differential evolution between red $L_*$ galaxies and their very massive counterparts, whose evolution seems to much more closely 
approximate that of a passively evolving galaxy population \footnote{For the purposes of this work, ``passive evolution'' refers to the
evolution of a galaxy in the absence of either mergers or ongoing star formation.}.
Note that, strictly speaking, a purely passively evolving galaxy population does not exist. 
As an example, mergers have been observed for individual massive elliptical galaxies (e.g. the Perseus A system).
An important question, however, is whether these processes are common enough to be significant in the average evolution of the galaxy population. 
One of the ultimate goal in galaxy evolution is to quantify the incidence of these processes.

\cite{Cool2008}, by computing the luminosity function (LF), found results consistent with passive evolution for the massive ($L>3L^*$) RS population at $z<1$. 
This conclusion was drawn from a combined sample of $\sim60,000$ LRGs at 
$0.1<z<0.4$ and $300$ LRGs at $0.6<z<1$. The small size of the high-redshift 
sample hinders, however, the interpretation of their results. Similar results for the very massive 
early-type population are found by \cite{Wake2006}, using both an LRG sample selected 
from SDSS data and the 2dF-SDSS LRG and QSO (2SLAQ) survey. This idea is reinforced by 
subsequent work from \cite{Maraston2013}, who adopted a refined version 
of the LRG passive template used in \cite{Cool2008} - described in \citealt{Maraston2009} -, to perform broad-band 
SED fitting on red galaxies from the BOSS DR9 dataset. \cite{Maraston2013} conclude that this template 
provides a good fit to BOSS galaxies redder than $g-i = 2.35$ at $0.4<z<0.7$, although there is only 
so much resolution that can be attained using broad-band SED fitting. The evolution in the clustering properties of LRGs from the BOSS 
samples also appear to support the idea of mergers having little effect in the evolution of these systems (see \citealt{Guo2013, Guo2014}).
Although passive evolution is just a simplified description, these results would 
indicate that this assumption well justified, within the uncertainties in the measurements.

Some other authors have reported results that suggest that mergers might still 
play a significant role, albeit small. \cite{Tojeiro2012} present a method
for identifying the progenitors of SDSS I/II LRGs within the BOSS sample. They conclude 
that the LRG population evolves slowly at late times, i.e., less than 2 per cent by merging from redshift $z \sim 0.55$
to $z \sim 0.1$, when the two samples are properly matched and weighted. \cite{Brown2007}, using data from the NOAO Deep 
Wide-Field and {\it{Spitzer}} IRAC Shallow surveys in the Bo\"otes field, find that the evolution of $4 L_*$ galaxies 
differs (although only slightly) from a model with negligible $z < 1$ star formation and no galaxy mergers.
From cross-correlation function measurements using $25,000$ LRGs at $0.16 < z < 0.36$ from 
the SDSS LRG sample, \cite{Masjedi2006} infer a low LRG-LRG merger rate, of $< 0.6 \times 10^4$ 
${\rm{Gyr}}^{-1}$ ${\rm{Gpc}}^{-3}$. \cite{Lidman2013} also found a major merger rate of $0.38\pm0.14$ 
mergers per Gyr at $z\sim1$, for brightest cluster galaxies (BCGs) using a sample of 18 distant galaxy 
clusters with over 600 spectroscopically confirmed cluster members between them (note 
that although many LRGs are BCGs, many LRGs are also field or group-member galaxies, so merger rates
between LRG and BCG populations are not strictly comparable). 

Interestingly, \cite{Bernardi2016} have recently reported some tension between LF/stellar mass
function (SMF) results and clustering results, although the authors also acknowledge that, again, a proper completeness 
correction for BOSS might alleviate this tension. \cite{Bernardi2016} claim that, while the 
evolution of the LF/SMF from BOSS to the SDSS is consistent with a passive evolution scenario, 
this appears not to be the case for the evolution of clustering. SDSS galaxies appear to be 
less strongly clustered than their BOSS counterparts, a result that seems to rule out the passive-evolution scenario, and, in fact, any minor merger scenarios 
where the rank ordering in stellar mass of the massive-galaxy population is preserved.

Constraining the effect of mergers in the evolution of massive RS galaxies is still, therefore, 
an open question, which has ramifications in other related fields.  
As an example, another piece of the puzzle has come from the study of the stellar populations of
individual early-type galaxies, since the seminal work of \cite{Tinsley1968}. 
It appears well established that the stars in the {\it{majority}} of the low-z, early-type galaxies were formed 
at $z\gtrsim2$, with little indications of star formation occurring since that time (\citealt{Bower1992}, \citealt{Ellis1997},
\citealt{Kodama1998}, \citealt{dePropris1999}, \citealt{Brough2002}, \citealt{McCarthy2004}, \citealt{Glazebrook2004}, \citealt{Holden2005},
\citealt{Wake2005}, \citealt{Bernardi2006}, \citealt{Jimenez2007}, \citealt{Thomas2005}). However, 
there appears to be a fraction that exhibit clear signs of recent star formation
(e.g., \citealt{Trager2000}, \citealt{Clemens2009}, \citealt{Schawinski2007}, \citealt{Balogh2005}). In this sense, 
indications of star formation are considerably more common among intermediate-to-low mass early-type galaxies 
and among early-type galaxies residing in lower-density environments (see, e.g., \citealt{Clemens2006} and 
\citealt{Thomas2010}).

The first goal of this paper is to provide an adequate forward-modeling framework for galaxy-evolution and clustering 
studies using BOSS. This framework is intended to characterize the high-mass end of the RS population at $z\sim 0.55$ 
(intrinsic distributions of galaxy properties, colour bimodality, LF) and to lay the foundations for a framework that can be extended to 
other DE surveys at other redshifts. We will concentrate on the photometric properties of the RS population; the inclusion of spectroscopic 
information is addressed in the complementary analysis presented in \cite{MonteroDorta2016}, in the context of the study of the $L-\sigma$ relation from BOSS.

The paper is organized as follows. Section~\ref{sec:overview} presents an overview of methods and 
motivations for this analysis. In Section~\ref{sec:data}, we describe 
the target selection for the BOSS CMASS sample and we discuss the observed CMASS distributions
in the colour-colour-magnitude space. Section~\ref{sec:stripe82} is devoted 
to briefly describing the procedure for modeling the covariance matrix for colours and magnitudes 
using Stripe 82 multi-epoch data. A detailed description of this aspect can be found in the Appendix. In Section~\ref{sec:expected_distribution} we discuss the 
expected shape for the massive end of the intrinsic distribution as predicted by stellar population synthesis models and 
provide some preliminary indications obtained from a simple histogram deconvolution method. 
In Section~\ref{sec:deconvolution} we describe in detail our analytical method for deconvolving the observed CMASS distribution. 
In Section~\ref{sec:results} we present the main results of this paper. Finally, 
in Section~\ref{sec:discussion} we discuss the significance and implications our results, in Section~\ref{sec:future}
we discuss future applications of our results and in Section~\ref{sec:conclusions} 
we briefly summarize the main conclusions of this study. Throughout this paper we adopt a cosmology 
with $\Omega_M=0.274$,  $\Omega_\Lambda=0.726$ and $H_0 = 100h$ km s$^{-1}$ Mpc$^{-1}$ with $h=0.70$ 
(WMAP7, \citealt{Komatsu2011}), and use AB magnitudes \citep{OkeGunn1983}.

\section{Overview of challenges, methods and motivations}
\label{sec:overview}

Despite its unprecedented statistical power, BOSS presents significant challenges, 
some of which are common to many future DE surveys (e.g., the Extended 
Baryon Oscillation Spectroscopic Survey, eBOSS; the Dark Energy Spectroscopic Instrument, DESI)
Measuring well-defined cosmological features, such as the BAO, results in survey requirements that are aimed at maximizing the global, statistical 
efficiency of the survey, at the cost of marginalizing the quality of individual-object measurements.
BOSS presents a combination of a complex target selection, 
large photometric errors and low S/N spectra (these last two issues being obviously related). In addition, model magnitudes, from which the colours used in the 
CMASS selection are built, are known to exhibit correlations between different bands, as 
the best-fit profile from the r band, convolved with the PSF in each band, 
is used as a matched aperture for all bands (\citealt{Strateva2001, Stoughton2002}). 

The above effects distort the observed color/magnitude distributions and 
hinder our ability to identify the latent intrinsic red/blue populations from BOSS data. This paper
is the first attempt to lay down an optimal framework for characterizing the galaxy 
population from BOSS and other future DE surveys. Our philosophy follows a forward-modeling approach based on the idea that 
the strategy followed by DE surveys should be reflected in the way we analyze the data: {\it{instead of 
individual galaxy properties our goal is to constrain distributions of galaxy properties}}. The main steps of our analysis are:

\begin{itemize}
	\item Present a method for the photometric deconvolution of the intrinsic colour-colour-magnitude distributions from the 
	observed BOSS CMASS distributions. We will proceed on the basis that the shape of the intrinsic distributions can be obtained directly from the data, without the need for making any strong assumptions based on stellar populations synthesis models.  The computed intrinsic distributions purely reflect the phenomenology of the colour-colour-magnitude diagrams.
	\item  Computation of completeness as a function of colors, magnitude and redshift in the BOSS CMASS sample. This is a key result 
	for future galaxy-evolution and galaxy-clustering studies.
 	\item Computation of the high-mass end of the RS LF within a redshift range around $z=0.55$.
	\item Computation of the best-fit passive-evolution model for the average evolution of the high-mass end of the RS LF.
	\item Evaluation and quantification of deviations from the best-fit passive-evolution model and possible systematics that affect this measurement.
	\item Quantification of the color evolution of the high-mass end of the RS.
\end{itemize}

The methodology and results presented in this paper constitute the basis for the scaling-relation analysis
presented in \cite{MonteroDorta2016}. In \cite{MonteroDorta2016} we use the intrinsic colour-magnitude distributions/completeness 
obtained from BOSS, in combination with velocity-dispersion likelihood functions to derive the high-mass end of 
the $L-\sigma$ relation at $z\sim0.55$. As part of the analysis we present a method for the photometric deconvolution 
of spectroscopic observable (named PDSO). 

In addition to the galaxy-evolution implications, understanding the BOSS data at the level of detail intended here has also important applications 
in other fields. In particular, this characterization of completeness and of the main statistical properties 
of the RS population will be used, in follow-up papers, in combination with N-body numerical simulations, to shed light 
into the halo-galaxy connection and the clustering properties of these systems in a fully consistent way. 

\section{Data: The BOSS CMASS sample} 
\label{sec:data}

In this paper we make use of spectroscopic and photometric data from the Tenth Data Release of the SDSS 
(DR10, \citealt{Ahn2013}). The SDSS DR10 is the third release within the SDSS-III, and the second 
release where BOSS data is included. The spectroscopic DR10 BOSS sample comprises a total of $927,844$ galaxy 
spectra and $535,995$ quasar spectra (an increase of almost a factor two as compared to the SDSS DR9, \citealt{Ahn2012}). The baseline 
imaging sample for the BOSS spectroscopic survey is the final SDSS imaging data set, which was released as 
part of the DR8 \citep{Aihara2011}, and contains imaging data from SDSS-I, -II, and -III\@.
These imaging programs provide five-band {\it{ugriz}} 
imaging over 7600 sq deg in the northern Galactic hemisphere and $\sim$ 3100 sq deg in the southern Galactic 
hemisphere, with a typical $50\%$ completeness limit for detection of point sources at $r = 22.5$. Refer to the
following references for technical information about the SDSS survey:
\cite{Fukugita1996} for a description of the SDSS {\it{ugriz}} photometric system; 
\cite{Gunn1998} and \cite{Gunn2006} for technical aspects of the SDSS camera and the SDSS telescope, respectively;
\cite{Smee2013} for information about the SDSS/BOSS spectrographs. 

The catalog that we use to compute the RS LF is the DR10 Large Scale Structure catalog (DR10 LSS). 
This catalog, which is thoroughly described in \cite{Anderson2014}, incorporates a detailed 
treatment of angular incompleteness (caused by fiber collisions and redshift failures) and of 
a variety of systematics that could potentially affect the target density of 
spectroscopically identified galaxies (e.g. stellar density, seeing, Galactic extinction, etcetera). The DR10 
LSS catalog includes galaxies from the SDSS Legacy 
Survey, which basically contains the SDSS-I survey and a small fraction of the SDSS-II survey. We
refer to \cite{Anderson2014} for further details about this catalog. 
The results presented in this paper, being limited primarily by systematic error, would not be expected to change significantly if the analysis were applied to the final, larger BOSS DR12 data set 
\citep{Alam2015}.

\begin{figure}
\begin{center}
\includegraphics[width=8cm]{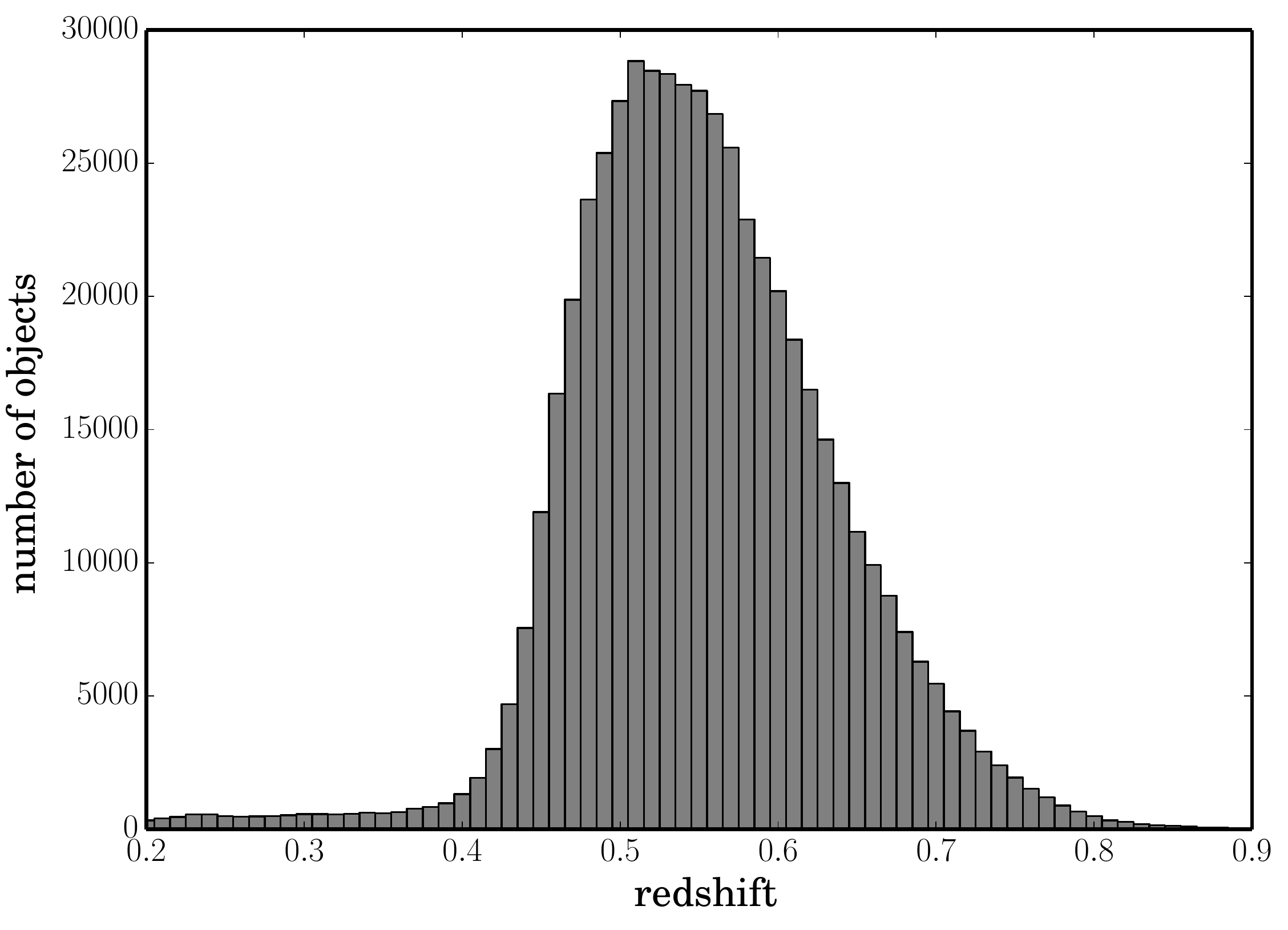}
\caption{Redshift distribution for the entire CMASS sample in bins of $\Delta z = 0.01$. 
The total number of unique CMASS galaxies with a good redshift estimate and with model and cmodel apparent magnitudes
and photometric errors in all g,r and i bands is 549,005.}
\label{fig:z_distribution}
\end{center}
\end{figure} 

\subsection{The CMASS galaxy target selection} 
\label{sec:BOSS}

The BOSS galaxy target selection is based on a similar strategy as that used by \cite{Eisenstein2001} to build the 
SDSS I-II LRG sample (for a detailed description of the BOSS selection see \citealt{Reid2016}). 
The selection was designed to produce two different galaxy samples: a low-redshift
sample called LOWZ and a high-redshift sample called CMASS. The LOWZ
sample contains LRGs within the redshift range $0.15<z<0.43$. 
The CMASS sample, which is the one that we use in this study, covers a nominal redshift range $0.43<z<0.70$, 
although it extends slightly beyond these limits. The acronym CMASS stands for ``constant mass", emphasizing the fact that the 
sample was designed to be approximately stellar mass-limited. This requisite naturally introduces a highly 
incomplete population of bluer galaxies that increases with redshift. 

The CMASS selection scheme consists of the following magnitude, colour and colour-magnitude cuts: \\

\begin{eqnarray} \displaystyle
17.5 < i_{cmod} < 19.9, \nonumber  \\
r_{mod} - i_{mod} < 2, \nonumber \\
i_{fiber2} < 21.5, \nonumber \\
d_\bot > 0.55, \nonumber\\
i_{cmod} < 19.86 + 1.6 \times (d_\bot - 0.8) 
\label{eq:selection}
\end{eqnarray}
 
\noindent where $d_{\bot} = (r_{mod} - i_{mod}) - (g_{mod} - r_{mod})/8$ and $i_{fiber2}$ is 
an i-band magnitude within a $2^{\prime\prime}$ diameter aperture. The subscript {\it{mod}} and {\it{cmod}} denote
model and cmodel magnitudes, respectively. Model magnitudes are more efficient at tracing galaxy colours, whereas the cmodel 
i-band flux is the best proxy for the galaxy flux. All colours 
quoted in this paper are {\it{model}} colours and all magnitudes {\it{cmodel}} magnitudes,
unless otherwise stated (hereafter we drop the subindices when we refer to 
these quantities). For more information on the BOSS selection 
refer to \cite{Reid2016}, \cite{Eisenstein2011} and \cite{Dawson2013}. \\

The process of generating the LSS catalog, as discussed in detail in \cite{Anderson2014}, is known to suffer from some degree of incompleteness, that 
mainly manifests itself in two forms: {\it{fiber collisions}} and {\it{redshift failures}}. Fibre collisions occur 
because of the finite size of the fiber plugs in the spectrographs. In BOSS, 2 fibers may not 
lie within $62^{\prime\prime}$ of one another on a given spectroscopic tile (see \citealt{Dawson2013} 
and \citealt{Reid2016} for more information) \footnote{Note that the fiber collision distance limit has
increased with respect to the SDSS I-II programs, from 
$55^{\prime\prime}$ to $62^{\prime\prime}$. As a reference, \cite{Blanton2003} estimate that 
fiber collisions account for approximately $6\%$ of all incompleteness in the SDSS survey.}.
With regard to redshift failures, the pipeline achieves an automated 
classification success rate of $98.70\%$ within the CMASS sample and confirms $95.4\%$ of unique CMASS targets as galaxies (from 
all spectra actually entering the pipeline, in the DR9). Note that the difference between these percentages 
arises from stars that pass the CMASS selection criteria (see \citealt{Bolton2012} for more information). 
These issues should have little correlation with colour or magnitude, which implies that 
this type of incompleteness simply translates into some small uncertainty in the determination of the normalization 
of the luminosity function.

In Figure~\ref{fig:z_distribution} we show the redshift distribution for the CMASS spectroscopic sample
in bins of $\Delta z = 0.01$. This is the redshift bin size that we use throughout this paper. The total number 
of unique CMASS galaxies with a good redshift estimate and with model and cmodel apparent magnitudes
and photometric errors in all g,r and i bands is 549,005. The mean value of the redshift distribution 
in Figure~\ref{fig:z_distribution} is  $0.532$ and its standard deviation $0.128$; 
approximately $\sim7.5 \%$ and $\sim4.5 \%$ of galaxies lie below and above the nominal low redshift 
and high redshift limits, i.e., $z=0.43$ and $z=0.70$, respectively.

\begin{figure}
\begin{center}
\includegraphics[scale=0.7]{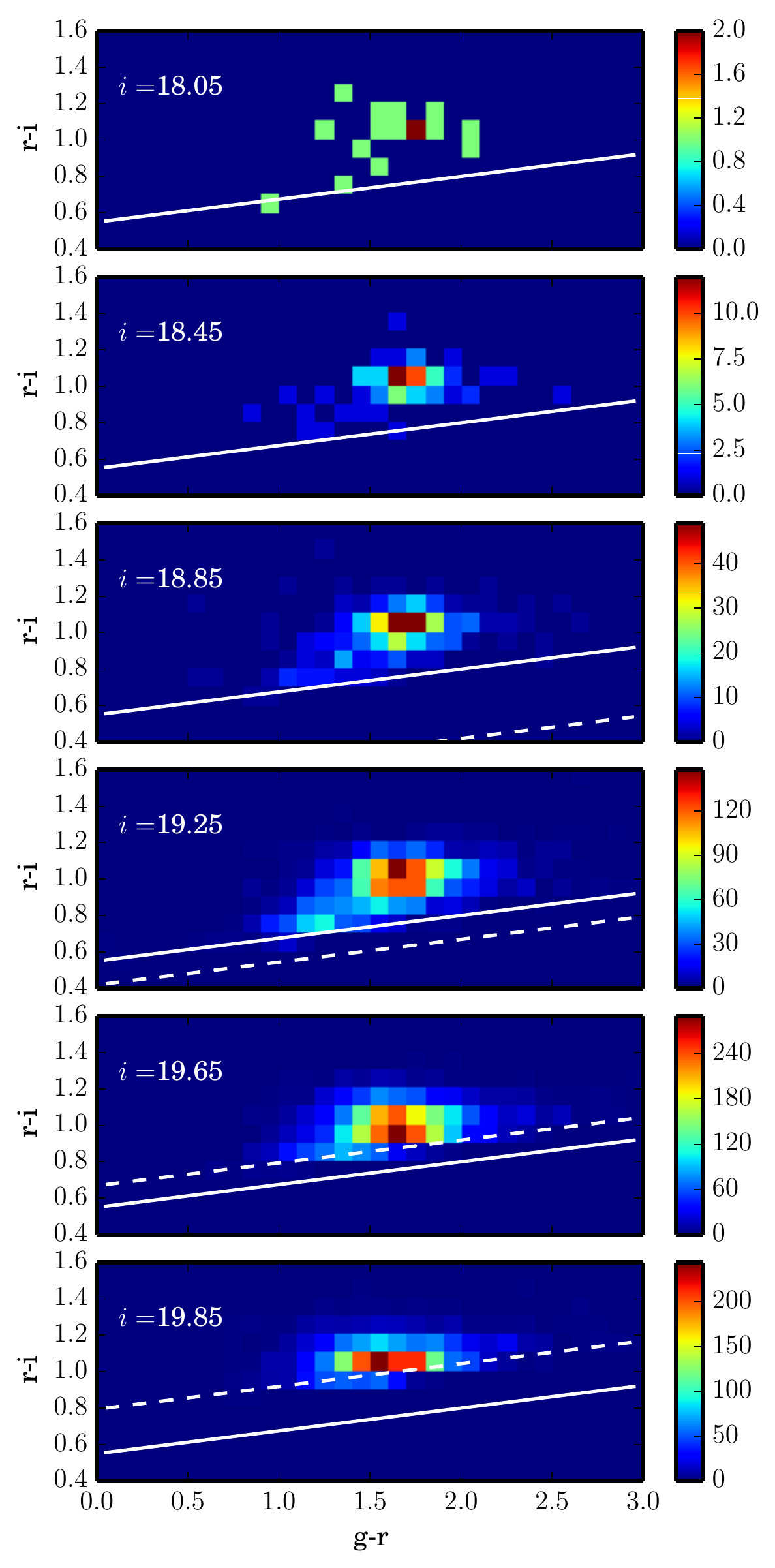}
\caption{Distribution of CMASS galaxies in a g-r vs. r-i colour-colour diagram for a redshift bin centered at $z=0.55$, with a width of $\Delta z=0.01$. In 
each panel, the solid and dashed lines represent the $d_\bot$ demarcation $d_\bot > 0.55$, and the sliding cut 
$i < 19.86 + 1.6 \times (d_\bot - 0.8)$, respectively. The observed CMASS distributions are considerably blurred by photometric
errors, especially along the g-r axis. The typical photometric error quoted in the catalog is $\sim 0.20$ for the 
g-r colour and $\sim 0.08$ for the r-i colour.}
\label{fig:ccd}
\end{center}
\end{figure} 

\begin{figure}
\begin{center}
\includegraphics[width=8cm]{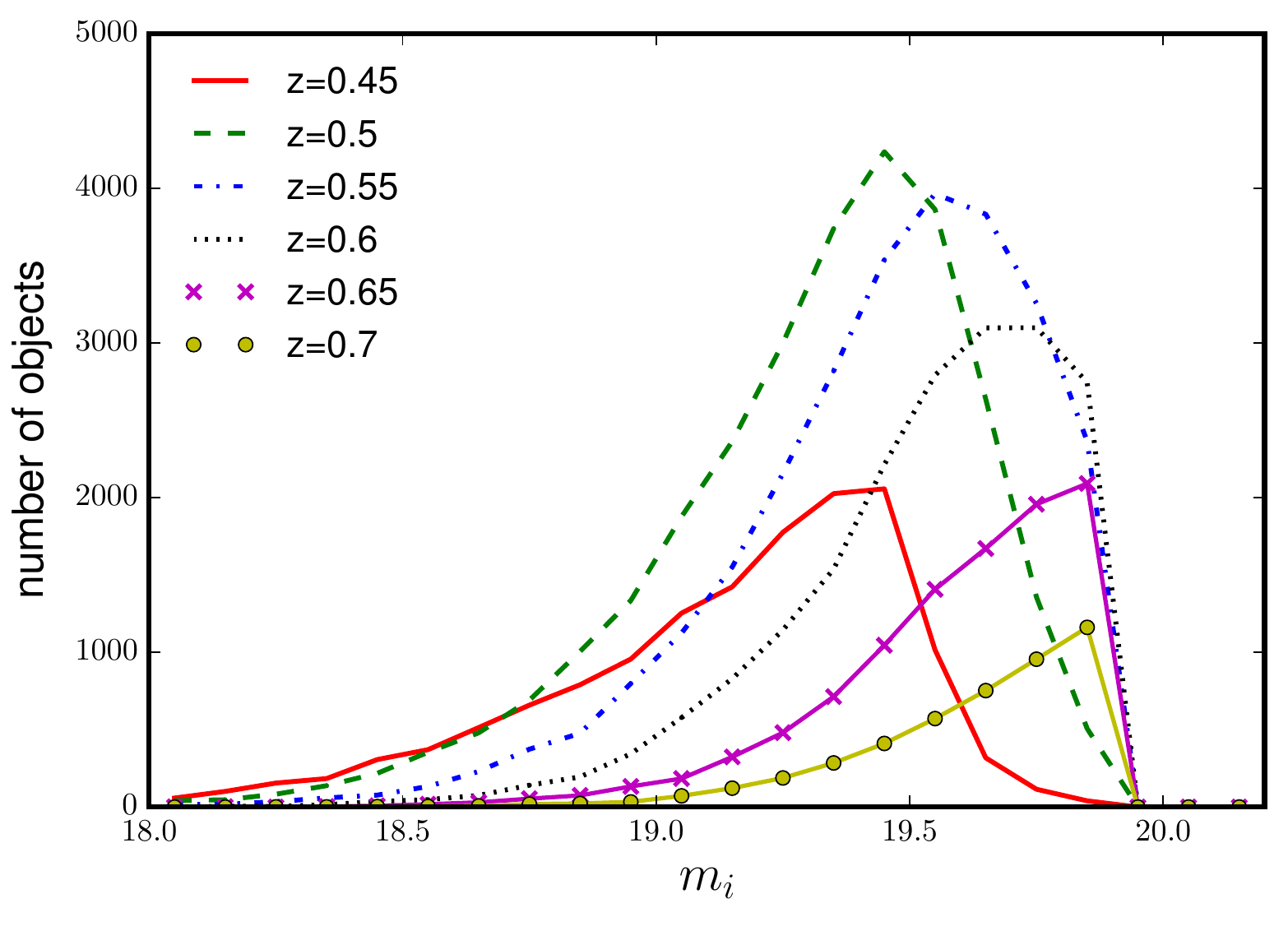}
\caption{CMASS number counts for different redshift slices, from $z=0.45$ to $z=0.70$. CMASS
number counts are strongly affected by selection effects and photometric errors. By determining 
the intrinsic number counts we can correct for this incompleteness.} 
\label{fig:nc}
\end{center}
\end{figure}

\subsection{CMASS observed distributions in colour-colour-magnitude space: the blurring effect.}
\label{sec:observed}

The CMASS observed distributions are not representative of the intrinsic, latent RS/blue cloud (BC) distributions. 
Figure~\ref{fig:ccd} presents in a g-r vs. r-i colour-colour diagram the distribution of CMASS galaxies in a redshift slice centered 
at $z=0.55$ ($\Delta z=0.01$), for different apparent magnitude bins. The $z=0.55$ bin is very close to the mean redshift of the CMASS 
sample (see Figure~\ref{fig:z_distribution}). Figure~\ref{fig:ccd} illustrates the type of broadened colour-colour distributions 
that we find in the BOSS galaxy data. The RS appears as an irregular blob, elongated mostly in the g-r direction 
due to the relatively large g-band errors, that 
is progressively more populated towards the fainter magnitude bins. Adjacent to the RS, in the g-i direction (diagonal 
in this diagram), is a population of bluer galaxies, which are not excluded by the CMASS selection. At $z=0.55$, these objects correspond, 
at least in observed space, to the upper, redder part of the BC and to the so-called green valley, the narrow region between the BC
and the RS. Colour bimodality is especially noticeable in the $i=19.25$ magnitude bin. 

The observed CMASS distribution is strongly affected by the $d_\bot$ constraint (solid line in Figure~\ref{fig:ccd}), that was designed to lie parallel 
to the locus of a passively evolving population of galaxies. This cut is supposed to isolate the red sequence, while allowing for a population of bluer 
galaxies. In practice, the $d_\bot$ demarcation is not fixed, but actually depends on magnitude.  It is 
determined by the combination of the $d_\bot$ cut and the sliding $d_\bot$ - i-band magnitude cut (dashed line in Figure~\ref{fig:ccd}). 
Rearranging in Equation~\ref{eq:selection}, the $d_\bot$ demarcation at a given apparent magnitude, $i$, is given by the following 2 inequations:

\begin{eqnarray} \displaystyle
d_\bot > 0.55, \nonumber\\
d_\bot > 0.8 + \frac{i-19.86}{1.6} 
\label{eq:selection}
\end{eqnarray}
 
\noindent which implies that the $d_\bot$ condition starts to be more restrictive at $i > 19.46$. It is also quite evident from Figure~\ref{fig:ccd} that the $r-i < 2$ cut 
has little impact on the CMASS distribution. 

The CMASS colour-colour diagram changes significantly when we examine 
different redshift slices, due to a combination of intrinsic evolution and the effect of redshift itself. 
In terms of the CMASS selection, the fraction of the RS and the BC accessible
increases with redshift. At $z\simeq0.45$, only the top of the RS is targeted, whereas at 
$z\simeq0.65$ the red sequence and the blue cloud are better covered (even though 
the effective magnitude range shrinks significantly). 

In Figure~\ref{fig:nc} we show the CMASS number counts, i.e. the distribution of CMASS galaxies in bins of i-band magnitude, for  six
different redshift slices, from $z=0.45$ to $z=0.70$. The shape of the number counts in Figure~\ref{fig:nc} 
is determined by several factors, including redshift, selection effects and 
photometric errors. The covariance matrix of photometric errors scatters objects around in the 3D space, including along the i-band axis.

\section{Modeling the covariance matrix with Stripe 82}
\label{sec:stripe82}

We use Stripe 82 multi-epoch data to model the blurring effect produced by photometric errors and correlations 
between different {\it{ugriz}} bands. Stripe 82 is an SDSS stripe along the celestial equator in the southern Galactic cap 
(covering a total of $\sim 270$ sq deg) that was observed multiple times, as many as $\sim 80$ times before the 
final release of the Stripe 82 database. These multi-epoch data are used to build our
model for the covariance matrix of the i-band magnitude, (g-r) color and (r-i) color (also represented here as ${X, Y, Z}$), a key 
element in our deconvolution procedure. 

A detailed description of the Stripe 82 data and the construction of 
our covariance matrix model can be found in the Appendix. Two important aspects of the modeling are, however, worth
highlighting here. Firstly, the model must account for the dependence of the elements of the covariance matrix 
on the location in color-color-magnitude space. Secondly, it must also accommodate possible overall inconsistencies in 
the photometric quality between the Stripe 82 and the SDSS footprint. To this purpose, we allow our covariance matrix model to be multiplied 
by a scale factor that could in principle have a small dependence on magnitude, but should be 
close to $1$. The value or functional form for this factor, that we call $\beta(i)$, will be empirically determined as part of the 
modeling of the intrinsic distributions.

\begin{figure}
\begin{center}
\includegraphics[scale=0.55]{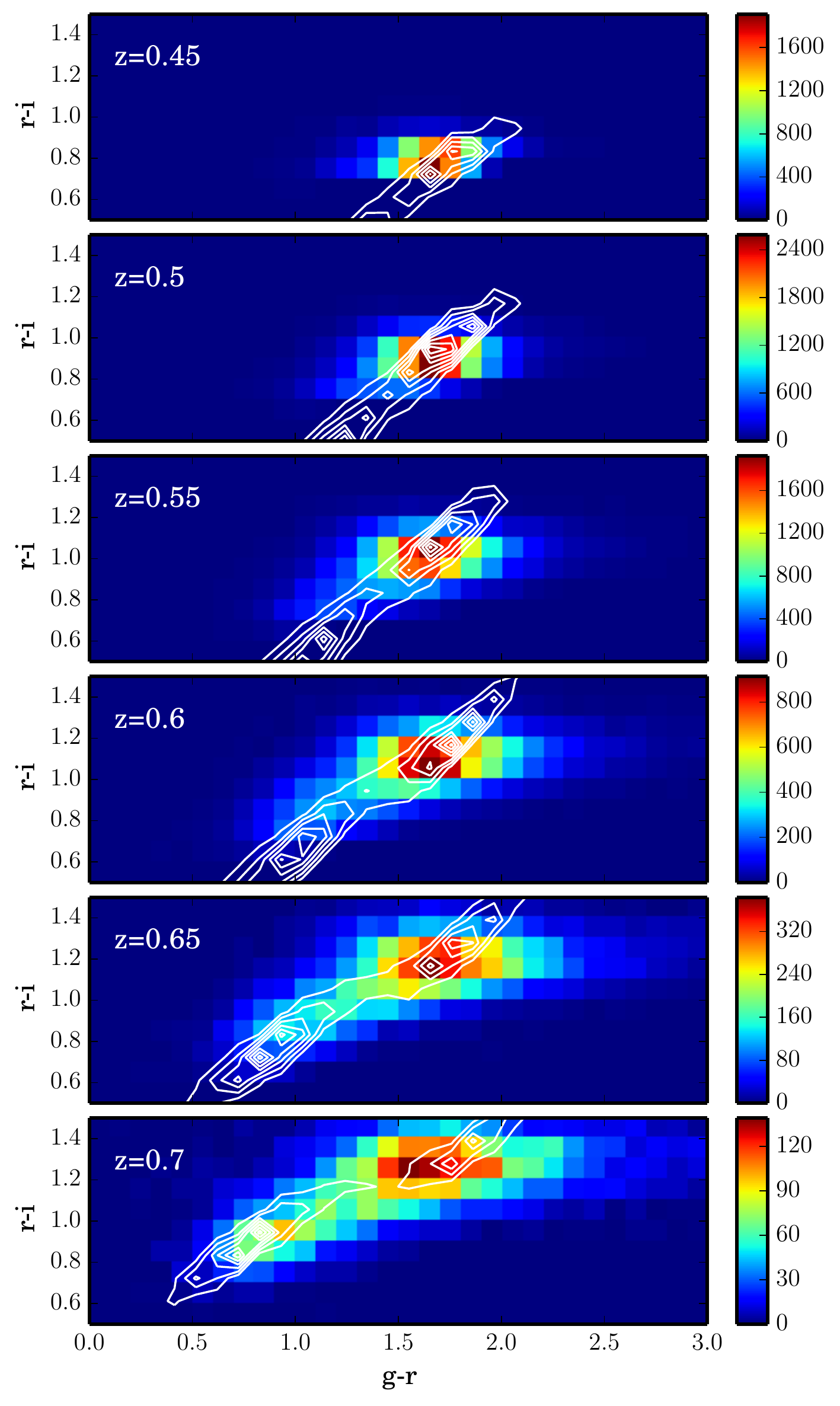}
\caption{The distribution of FSPS models (white contours) in a g-r vs. r-i colour-colour diagram, for six different redshift slices from $z=0.45$ to $z=0.70$. FSPS 
models are taken from the Granada FSPS galaxy product, which is part of the SDSS DR10 \citep{Ahn2013}. These FSPS models cover a wide 
range of metallicities, SFHs and dust attenuation levels. The colour map shows the distribution of CMASS galaxies 
in the corresponding redshift slice. The typical photometric error quoted in the catalog is $\sim 0.20$ for the 
g-r colour and $\sim 0.08$ for the r-i colour.} 
\label{fig:datavsfsps}
\end{center}
\end{figure}

\section{Hints on the shape of the intrinsic colour-colour-magnitude distribution}
\label{sec:expected_distribution}

Although stellar population synthesis (SPS) models are known to suffer from 
major uncertainties (see \citealt{Conroy2009} for a discussion), they provide
preliminary information on the typical intrinsic location in the colour-colour plane of physically-plausible galaxies in the 
absence of photometric errors. In order to explore the predictions from these models, we use the Granada FSPS galaxy product (Montero-Dorta et al., in preparation) 
based on the publicly available ``Flexible Stellar Population Synthesis'' FSPS code of \citet{Conroy2009}. The Granada FSPS product, 
which is part of the SDSS DR10 \citep{Ahn2013}, provides spectro-photometric stellar masses, ages, specific star formation 
rates, and other stellar population properties, along with corresponding errors, for the entire DR10 galaxy sample. Here we use the same grid of 
CSP models with varying star formation history (based on simple $\tau$-models), metallicity and dust attenuation used to generate the Granada FSPS 
product. This extensive grid of $84,000$ models (in its complete version, before applying any priors) was designed to adequately cover the CMASS parameter space.

Figure \ref{fig:datavsfsps} displays, in white contours, the distribution of FSPS models in a g-r vs. r-i colour-colour diagram, for six 
redshift slices from $z=0.45$ to $z=0.70$. The only prior imposed here is that the look-back formation time 
of the galaxy cannot exceed the age of the Universe at the corresponding redshift. As mentioned above, 
models expand a wide range of stellar population properties, so the width of the model distribution should be considered
an upper-limit prediction for the intrinsic distribution. In the background of each panel, we show the distribution of CMASS galaxies 
in the corresponding redshift slice. The FSPS models align extremely well with the data at every redshift. As expected, models expand a narrow intrinsic distribution 
in colour space, which suggests that the observed CMASS distribution is largely broadened by photometric errors. 
Interestingly, colour bimodality appears naturally in the grid. In addition, the RS and the BC form 
a continuous sequence at redshifts $z \lesssim 0.55$ in the models. This sequence clearly breaks 
at $z \gtrsim 0.60$, so that both components are no longer aligned (note that a lack of alignment 
of the RS with the BC has been reported before using optical - IR colour;
a similar result was found by \citealt{Brown2014} in the $z\sim0$ u-g-r colour-colour diagrams 
for nearby galaxies). 

The distribution of models in Figure \ref{fig:datavsfsps} has not been calibrated with observations,
so it simply provides a sense for the intrinsic loci of RS objects. Tighter constraints on the 
shape of the intrinsic RS distribution are obtained from a simple histogram deconvolution of the CMASS observed 
distributions. Although this method is not accurate enough in the context of our analysis, as binning 
effects can propagate into the computation of the LF, it does provide information on the general shape of the 
intrinsic distributions. What we find is that only an extremely compact ($< 0.1$ mag) RS 
can be consistent with our photometric error model. In addition, a declining colour-magnitude 
relation (for the centroid of the RS distribution) is necessary to fit the observations. Another important 
piece of information that we obtain from this simple analysis is that the effect of the covariance 
matrix obtained from our error model must be slightly reduced in order to further achieve a good 
agreement with the observations (i.e. $\beta \sim 0.92$).

Our modeling of the intrinsic distributions is not based on SPS models, nor it uses a histogram deconvolution method. The knowledge 
gained from these approaches is however transmitted into our deconvolution method in the form of 
constraints that reduce the dimensionality of our parameter space.

\section{A method for the analytical deconvolution of intrinsic distributions of photometric properties from BOSS}
\label{sec:deconvolution}

\begin{table*}
\begin{center}
\small{
\begin{tabular}{cc}
\hline
\hline
      PDF parameter &    Description    \\
\hline
      $m_*$   & Characteristic apparent magnitude for intrinsic number counts  \\

      $\alpha$   &  {\raggedleft{}Faint-end slope of the LF}    \\

      $Y_0$    &   Zero point for the linear dependence of the (g-r) component of the centroid with magnitude \\

      $Y_1$  &  Slope for the linear dependence of the (g-r) component of the centroid with magnitude \\

      $Z_0$ &     Zero point for the linear dependence of the (r-i) component of the centroid with magnitude \\
      
      $Z_1$ & Slope for the linear dependence of the (r-i) component of the centroid with magnitude \\
      
      $\sigma$  &  Intrinsic scatter on the colour-colour plane \\
      
      $\theta$  & Position angle of the Gaussian Function on the colour-colour plane w.r.t. the g-r axis \\
      
      $q$  &    Inverse of the ellipticity of the Gaussian Function on the colour-colour plane \\
      
      $f_{blue}$  & Observed fraction of blue objects in the CMASS sample     \\

      $\beta$ &  Error scale factor for the covariance matrix model    \\         
\hline
\hline
\end{tabular}}
\end{center}
\caption{Fitting parameters of the PDF for each Gaussian function ($m_*$, $\alpha$, $Y_0$, $Y_1$, $Z_0$, $Z_1$, $\sigma$, $\theta$, $q$) along with $f_{blue}$ and  $\beta$}
\label{table:parameters}
\end{table*}

In this section, we describe a forward-modeling method for the deconvolution of the 
CMASS intrinsic colour-magnitud distributions. The method is based on an unbinned maximum likelihood method (UML).  
The key idea for the UML is that the underlying statistical quantity that we optimize, 
the likelihood $\mathscr{L}$ (or $log \mathscr{L}$), does not require any artificial 
binning beyond the one produced by the measurement instruments. Although, for the sake of 
clarity, we will describe our deconvolution method in the context of BOSS data, our method can be 
naturally extended to future DE surveys like eBOSS or DESI. 

The UML method is based on constructing the predicted probability density function (PDF) 
for the data as a function of the parameters of interest. Once we have an analytical parametric 
expression for this PDF, the likelihood of a particular set of values for the parameters, $\theta$, given the data, $\vec{d}$, can 
be obtained. This likelihood is equal to the probability, $P$, of measuring the data 
given that set of values, which is equal to the product of all individual probabilities, $p_i$. Namely:

\begin{equation}
\mathscr{L}(\theta|\{\vec{d}\}) = P(\{\vec{d}\}|\theta) = \prod_i p_i(\{\vec{d_i}\}|\theta)
\label{eq:likelihood}
\end{equation}

To find the best-fit parameters it is common to optimize the 
logarithm of the likelihood, $log \mathscr{L}(\theta|\{\vec{d}\})$, instead of the likelihood itself.

The observed PDF for the distribution of CMASS objects in the colour-colour-magnitude space 
is shaped by contributions from the intrinsic distribution, the blurring effect of photometric errors
and the CMASS selection. For the sake of simplicity, we will first restrict the analysis to the RS.
Results from a simple histogram deconvolution method suggest that the RS can be described with sufficient accuracy in observed space
by a Gaussian function representing the colour-colour component and a Schechter function 
in apparent magnitude accounting for the i-band magnitude dependence. This information is 
incorporated in our analytical modeling of the PDF. 

In this paper, we focus on the deconvolution of photometric quantities (colour-magnitude distributions). 
Many galaxy-evolution science cases using DE surveys will involve distributions of spectroscopic quantities. The 
photometric deconvolution of spectroscopic observables (e.g. velocity dispersion) is addressed in \cite{MonteroDorta2016}.

\subsection{The CMASS probability density distribution}
\label{sec:PDF}

In order to build the observed CMASS PDF, we start by expanding the Schechter function 
(that accounts for the intrinsic number counts) in terms of a non-negative sequence of Gaussians along the i-band magnitude 
axis:

\begin{eqnarray} \displaystyle
n_{sch}(X,\{\tilde{\phi_*},m_*, \alpha\}) = \sum_j c_j \mathcal{G}_j
\label{eq:expansion}
\end{eqnarray}

\noindent where $n_{sch}$ represents the intrinsic number counts (i.e., the intrinsic number of objects at each magnitude bin), $\{\tilde{\phi_*},m_*, \alpha\}$ 
are the apparent-magnitude counterparts of the Schechter LF parameters, $c_j$ is coefficient $j$ of the expansion and $\mathcal{G}_j$ is the corresponding 
Gaussian function. By expanding the Schechter function as a series of Gaussians, we are able to implement an analytic convolution of the luminosity function with 
the Gaussian colour-colour-magnitude error model. The Schechter Function can be expressed as:
  
\begin{eqnarray} \displaystyle
n_{sch}(X,\{\tilde{\phi_*},m_*, \alpha\}) = 0.4 {\rm log}(10) \tilde{\phi_*} \left[10^{0.4(m_*-X)(\alpha+1)}\right] \times  \nonumber \\
{\rm exp}\left(-10^{0.4(m_*-X)}\right) 
\label{eq:schechter}
\end{eqnarray}

Here, we assume that $\alpha = -1$ and $\tilde{\phi_*} = 1$ object, for the sake of simplicity. The magnitude interval considered is 
$17 < X < 20.5$, which exceeds by half a magnitude the CMASS limits at both ends, in order to avoid boundary problems. For convenience, 
we choose to use 22 Gaussian functions to completely cover this interval, with equal width $\sigma_i = 0.25$ mag. 

The above expansion can be generalized to account for the colour component at each apparent magnitude bin. 
In particular, we can model the PDF at a given magnitude bin $k$ as a 
3D Gaussian of i-band magnitude (X), g-r colour (Y) and r-i colour (Z). In a compact notation, this function can be written as follows:

\begin{eqnarray} \displaystyle
\mathcal{P}(V|\theta) = \frac{c_k}{(2\pi |U|)^\frac{3}{2}} {\rm exp}\left[-\frac{1}{2} (V-V_0)^T U^{-1} (V-V_0)\right] \nonumber \\
\label{eq:PDF}
\end{eqnarray}
 
 \noindent where $V$ is a vector of coordinates in the colour-colour-magnitude space, i.e., $V = \{X,Y,Z\}$, and $V_c$ is the vector of coordinates 
 at $\{X_c, Y_c, Z_c\}$, which corresponds to the center of Gaussian $k$ along the i-band axis and the
 centroid of the RS on the colour-colour plane, respectively. All the information regarding the 
 shape of the PDF is contained in matrix $U$, including both the contribution from the intrinsic distribution and 
 the covariance matrix for the photometric errors, $C$. It can be shown that if the intrinsic model is assumed to be 
 a 3D Guassian where $\sigma_i$ is the intrinsic size along the magnitude axis and  $\sigma_c$ is the intrinsic scatter on the 
 colour-colour plane, matrix $U$ can be expressed as:

\[ 
U =  C +
  \begin{bmatrix} 
  \sigma_i^2 & 0 & 0 \\ 
  0 &  \sigma_c^2(q^2 a^2 + b^2) & \sigma_c^2 a b (1 - q^2 )\\ 
  0 & \sigma_c^2 a b (1 - q^2 )  & \sigma_c^2(q^2 b^2 + a^2)   
  \end{bmatrix} 
\]

\noindent where $\theta$ is the position angle of the Gaussian on the colour-colour plane w.r.t. the X axis, $a = \sin(\theta)$,
$b = \cos(\theta)$ and {\it{q}} is the inverse of the ellipticity. The second term in the above equation 
accounts for the contribution of the intrinsic distribution on the covariance matrix for the PDF - hereafter 
$C_{int}$. 

As mentioned before, we need to account for the possible shift of the RS 
centroid with magnitude on the colour-colour plane. To this end, we model $\{Y_c, Z_c\}$ as:

\begin{eqnarray} \displaystyle
Y_c = Y_0 + Y_1 (X-X_{ref})  \nonumber \\
Z_c = Z_0 + Z_1 (X-X_{ref})
\label{eq:schechter}
\end{eqnarray}

\noindent where $X_{ref}$ is a reference magnitude, for which we choose $X_{ref} = 19$. 

Expression~\ref{eq:PDF} is still not a fully-representative model for the observed CMASS PDF, 
because the CMASS selection has not yet been taken into account. This step is done trivially 
by applying the CMASS selection matrix, $S_{CMASS}$. Namely:

\begin{equation}
\mathcal{P}_{CMASS}(V|\theta) = \mathcal{P}(V|\theta) S_{CMASS} 
\label{eq:cmass_sel}
\end{equation}

A single-Gaussian model is a good approximation for the 
RS, but not for the entire CMASS sample, which contains a fraction 
of blue objects that increases with redshift. In order to take this population into account, 
we use a double-Gaussian model for the CMASS PDF:

\begin{eqnarray} \displaystyle
\mathcal{P}_{CMASS}(V|\theta) = f_{blue} \mathcal{P}_{CMASS}^{BC}(V|\theta^{(BC)})  \nonumber \\
+ (1-f_{blue})\mathcal{P}_{CMASS}^{RS}(V|\theta^{(RS)}) 
\label{eq:complete_PDF}
\end{eqnarray}

\noindent where $\mathcal{P}_{CMASS}^{BC}$ is the PDF for the BC, $\mathcal{P}_{CMASS}^{RS}$ is the PDF 
for the RS and $f_{blue}$ is the observed fraction of blue objects in the 
sample (blue objects meaning non-RS objects). With this extension,  
the total number of fitting parameters is 19 (2$\times$9+1). In the following sections we discuss 
possible priors that can reduce the number of parameters in our PDF model. 

Even though we use the standard terminology ``blue cloud/BC" here, this distribution must not 
be assumed to correspond to any previous definitions of blue cloud or blue objects. Our BC distribution
is a second component that we need in order to fit the observed distributions and that is statistically  
separable from the prominent red sequence (see a similar approach in \citealt{Taylor2015}).

Table~\ref{table:parameters} lists all parameters of the PDF for each individual Gaussian function, 
along with the parameters accounting for the fraction of blue objects, $f_{blue}$. In addition, 
we include $\beta$, the error scale factor for the covariance matrix $C$.

\subsection{Computation of best-fit parameters}
\label{sec:computation}

The determination of best-fit parameters is performed by simply minimizing the $-{\rm log} \mathscr{L}$. The 
likelihood $\mathscr{L}$ for a given set of parameters can be obtained at each redshift 
slice by evaluating our observed CMASS PDF at the position of each object in our 3D space, 
and summing the probabilities. 

The above optimization method provides a best-fit intrinsic distribution at each 
redshift slice. As part of this process, it provides $m_*$ which, as we demonstrate in subsequent sections, can be 
trivially transformed into $M_*$, the characteristic absolute magnitude of the LF.

Our method also incorporates an easy way to compute $\phi_*$, the normalization 
of the RS LF. We have assumed for simplicity that  $\tilde{\phi_*}$, the
normalization of the Schechter function that represent the RS intrinsic distribution, is equal to 1. 
The predicted fraction of objects that are selected for the CMASS sample, after convolving 
with the error model, is given by:

\begin{equation}
n_{pred} = \sum_i \mathcal{P}_{CMASS}^{RS}(V_i|\theta) \Delta X \Delta Y \Delta Z 
\label{eq:n_int}
\end{equation}

\noindent where this computation must be performed prior to any normalization. If 
the total number of real CMASS objects detected within the corresponding redshift slice
is given by $n_{obs}$, the value of $\phi_*$ for the RS LF can be calculated by:

\begin{equation}
\phi_*^{RS} = \frac{(1-f_{blue})\frac{n_{obs}}{n_{pred}}}{V_{max}}
\label{eq:phi_star}
\end{equation}

\noindent where $(1-f_{blue})\times\frac{n_{obs}}{n_{pred}}$ is the predicted number of intrinsic RS objects and 
$V_{max}$ is the maximum volume in the corresponding redshift slice. A similar
procedure can be used to obtain $\phi_*^{BC}$.

In the case of CMASS sample, as we discuss in the next sections, the Schechter parameters $\phi_*$ and $M_*$
must be exclusively considered the parameters of an analytic functional form for the computed LF. A physical 
interpretation of these parameters is misleading, given that the CMASS sample probes only 
the very bright end of the LF, more than one magnitude brighter than $M_*$.

\begin{figure}
\begin{center}
\includegraphics[scale=0.6]{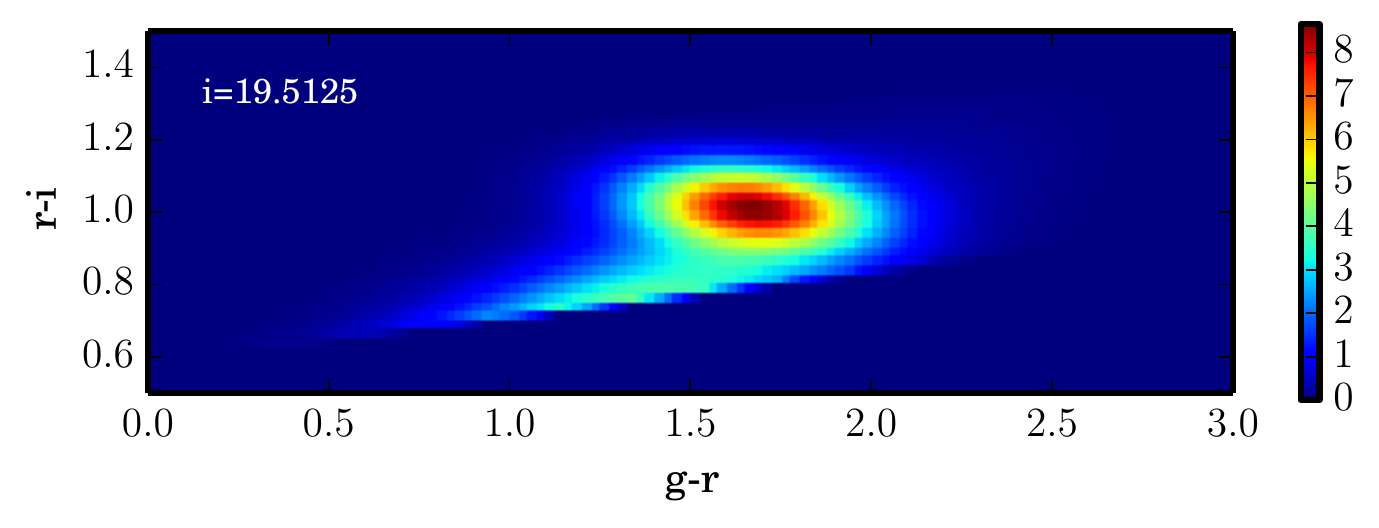}
\caption{Best-fit model for the CMASS PDF in a redshift slice centered at $z=0.55$ ($\Delta z = 0.01$) 
obtained using our analytical deconvolution method. The analytical form of the CMASS PDF contains 
both the contribution of the intrinsic distribution and the effect of photometric blurring and selection 
effects.}
\label{fig:grid_pdf}
\end{center}
\end{figure}

\subsection{Adopted forms for the intrinsic RS and BC distributions}
\label{sec:priors}

We use our histogram deconvolution method 
to place constraints on the analytical model for the CMASS PDF described above.
This helps us reduce the dimensionality of our parameter space.
The following information is incorporated into our model:

\begin{itemize}
	\item The RS is consistent with a single point on the 
	colour-colour plane at fixed magnitude, with all the observed scatter being due to photometric errors, i.e.,
	$q^{RS}=0$, $\sigma^{RS}=0$. 
			
	\item The location of the RS in the 
	colour-colour plane must have a slight dependence on magnitude.
		
	\item Our error model alone cannot account completely for the scatter seen on the blue side, even though a simple 1-dimensional Gaussian, 
	a line in the colour-colour plane, is certainly not a bad approximation. 
		
	\item Due to the severe incompleteness affecting the blue side, the BC
	must only be modeled for the sake of subtracting its contribution on the red side.
	
	\item  Allowing too much freedom to the model when finding the location of the BC centroid might 
	result in unphysical solutions. 
	
\end{itemize}

\begin{figure*}
\begin{center}
\includegraphics[scale=0.95]{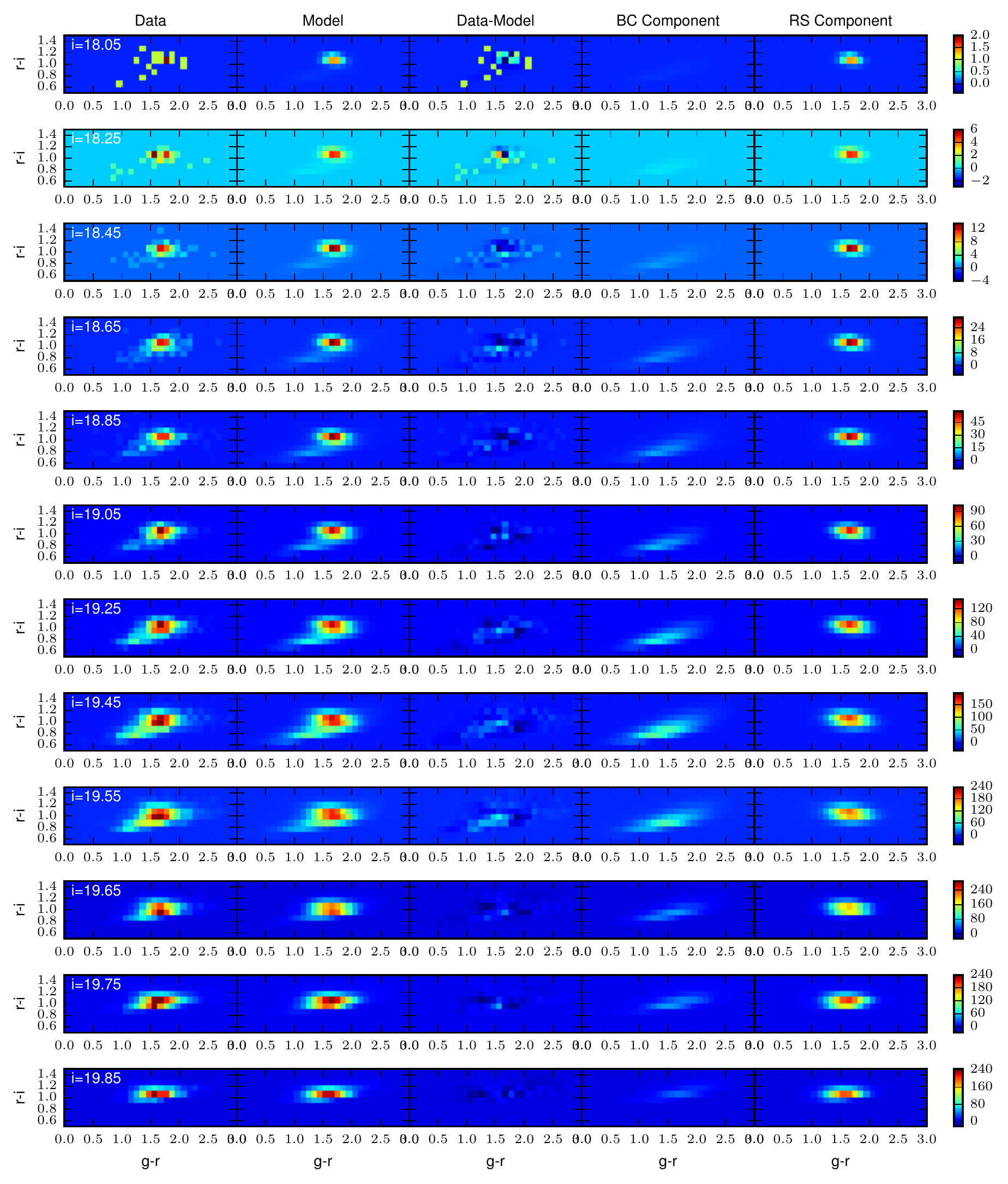}
\caption{Best-fit model in observed space as compared with the data, in a redshift slice of width $\Delta z=0.01$ centered at $z=0.55$. 
From left to right: data, model, data-model residuals, model for the red sequence (RS) component and model for the blue cloud (BC) component, respectively, in 
g-r vs r-i colour-colour diagrams for $12$ different magnitude bins. The bin size is $0.1$ in all three 
quantities. The same colour code applies to all distributions in the same panel, but colour codes vary for different 
panels. The typical photometric error quoted in the catalog is $\sim 0.20$ for the 
g-r colour and $\sim 0.08$ for the r-i colour. Our CMASS PDF reproduces, to a high level of accuracy, the shape and main characteristics of the observed CMASS
distributions and also accounts for its magnitude dependence.}
\label{fig:models055_adm}
\end{center}
\end{figure*}

\begin{figure*}
\begin{center}
\includegraphics[scale=0.42]{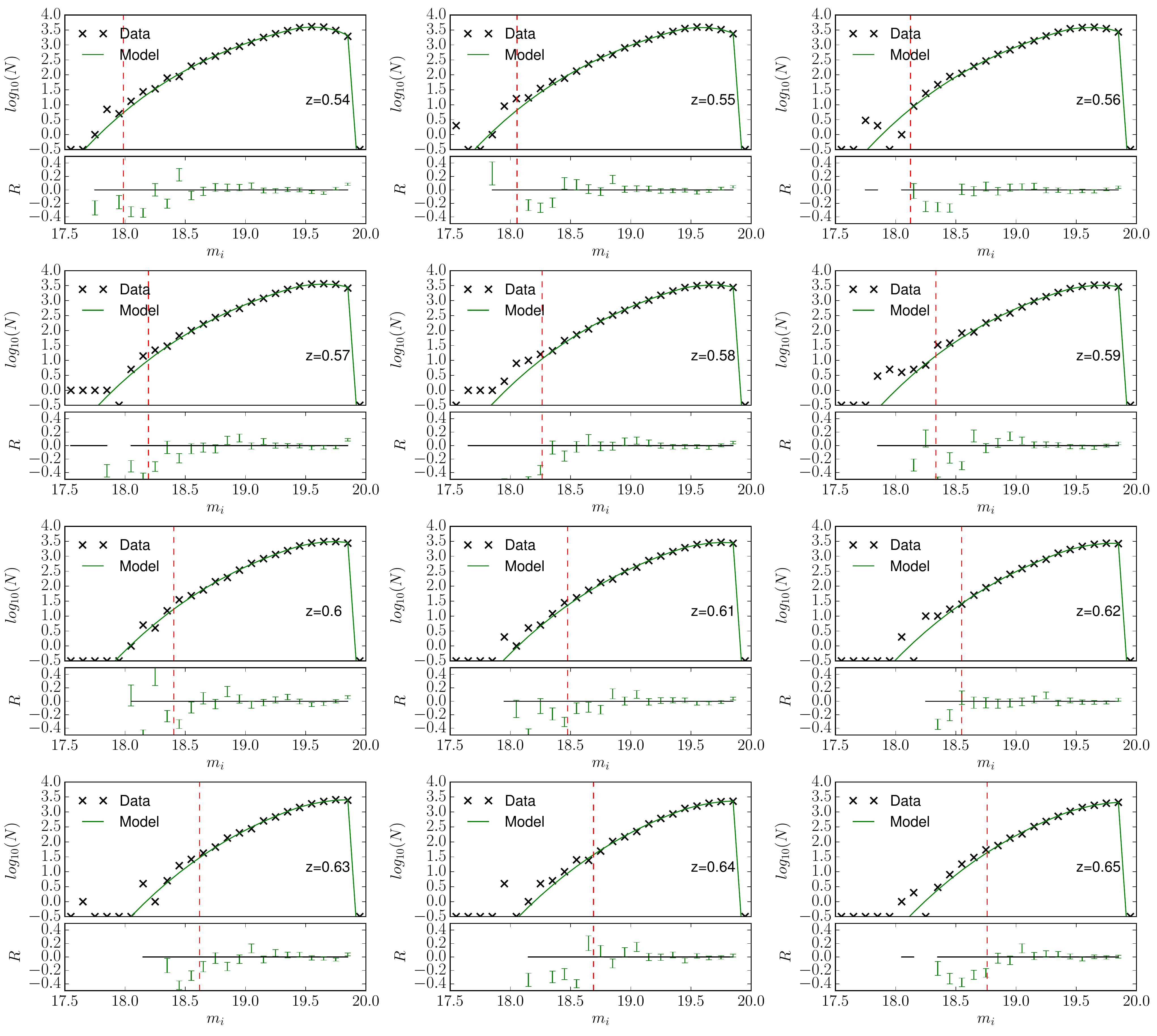}
\caption{The i-band number counts for both the data and the model at 12 redshift slices from 
$z=0.51$ to $z=0.70$, illustrating that the agreement between the 
model and the data is quite good across the redshift range considered (see text). Each upper panel shows the logarithm of the number of 
galaxies in each magnitude bin, $N(m)$, as a function of magnitude for the data (black crosses) and the model in 
observed space (green solid line). The subplots show the fractional residuals, where the errors in the model 
are obtained from the scatter on the Schechter best-fit parameters. The vertical dashed line is the apparent magnitude 
corresponding to a K-corrected absolute magnitude of $^{0.55}M_i =  -24$.}
\label{fig:nc_adm_z}
\end{center}
\end{figure*}

With the above considerations, the intrinsic component of the covariance matrix in 
Equation~\ref{eq:PDF}, for the single-point RS, can be simply expressed as:

\[ 
C_{int}^{RS} =  
  \begin{bmatrix} 
  \sigma_i^2 & 0 & 0 \\ 
  0 & 0 & 0\\ 
  0 & 0  & 0   
  \end{bmatrix} 
\]

\noindent which is equivalent to a Delta function on the colour-colour plane, with position given 
by $Y_0^{RS}$, $Y_1^{RS}$, $Z_0^{RS}$, $Z_1^{RS}$.

As for the BC component, we keep $C_{int}$ in its general form, namely:

\[ 
C_{int}^{BC} =  
  \begin{bmatrix} 
  \sigma_i^2 & 0 & 0 \\ 
  0 &  \sigma_c^2(q^2 a^2 + b^2) & \sigma_c^2 a b (1 - q^2 )\\ 
  0 & \sigma_c^2 a b (1 - q^2 )  & \sigma_c^2(q^2 b^2 + a^2)   
  \end{bmatrix} 
\]

\noindent but we impose $Y_1^{BC}$ = 0 and $Z_1^{BS}$ = 0. We also 
include the condition $Y_0^{BC}$ + $Z_0^{BC}$ = 1.7, independently 
of redshift. By doing so, we can fix the centroid of the blue cloud at $g-i = 1.7$, ensuring 
an adequate BC/RS separation. This value was chosen by examining 
the distribution of FSPS models (Figure~\ref{fig:datavsfsps}) and  
the distribution of objects in the PRIsm MUlti-object Survey (PRIMUS, \citealt{Coil2011, Cool2013}), which is not 
affected by a colour cut. We have also checked, and this is visible in 
Figure~\ref{fig:datavsfsps}, that the observed centroid of the blue cloud is stable
as we move across redshift, due to the typical shape of the SED for blue galaxies, which lack 
a $4000$ \AA~ break. Note that the BC here acts as a background component
that is necessary to fit the observations.

\subsection{Parameter uncertainty estimation}
\label{sec:uncertainty}

In order to estimate the uncertainty in the fitted parameters, we start by 
estimating the statistical errors, by means of a bootstrap analysis.
We generate 250 bootstrap samples by randomly drawing objects from the 
entire CMASS sample, allowing for objects to be repeated. Bootstrap samples are 
chosen to have the same size of the CMASS sample.  

In principle, a single value of $\beta$ should hold across all redshift bins. In practice, if
we fit for $\beta$ as a free parameter in each redshift bin, the range of fitted values
across redshift exceeds the amount that would be expected based the on bootstrap-estimated
statistical errors on the individual $\beta$ estimates. This flexibility in $\beta$
can absorb residual un-modeled redshift-dependent effects in the deconvolution.

In order to quantify the systematic effect
of this $\beta$ uncertainty upon our estimated luminosity-function parameters, we
re-fit our model in each redshift bin with two fixed $\beta$ values chosen to approximately
bracket the distribution of free-$\beta$ fits across redshift. The half-width of the range in fitted
LF parameter from this procedure is taken as an estimate of the systematic error in those
parameters, and is added in quadrature with the bootstrap-estimated statistical errors
to give our overall LF uncertainty estimate. Note that the systematic term is in all cases
significantly larger than the statistical term, due to the extremely large number of
galaxies in each redshift slice.

\section{Results}
\label{sec:results}
\subsection{Algorithm performance}
\label{sec:performance}

The performance of the analytical deconvolution method is excellent in terms of reproducing the 
observed distributions for redshift slices within 
the redshift interval $0.53\lesssim z \lesssim 0.65$.  Below $z\simeq0.53$, results become progressively noisier
as the CMASS RS is more and more incomplete in colour. 
Above $z \simeq 0.65$, our method converges and 
provides acceptable residuals (up to $z=0.70$), but the effective magnitude range is so small that
some of our results may become rather unconstrained (see the LF section below).  

Figure~\ref{fig:grid_pdf} displays our best-fit model for the CMASS PDF 
in the redshift slice centered at $z=0.55$, for an apparent magnitude $i=19.5$. 
As described before, this PDF provides the probability density 
of finding a galaxy at any given point in the colour-colour plane, for our best-fit error-deconvolved 
intrinsic distribution, our Stripe 82 error model and the CMASS selection function. Figure~\ref{fig:grid_pdf}
shows that the CMASS PDF is consistent, generally speaking, with
the observed CMASS distribution. In order to illustrate this more clearly, we have 
generated a mock distribution assuming this best-fit CMASS PDF, with 
the same number of objects as those passing the CMASS selection 
in the $z=0.55$ redshift slice. Figure~\ref{fig:models055_adm} compares 
these two distributions, which are presented 
in bins of magnitude ($\Delta i = 0.1$) and colours ($\Delta g-r = 0.1$, $\Delta r-i = 0.1$). 
In particular, we show the data, the model, the data--model residuals and the RS
and BC components, respectively. 

A visual inspection of the residuals in Figure~\ref{fig:models055_adm} suggests that our CMASS
PDF reproduces, to a high level of accuracy, the shape and main characteristics of the observed CMASS
distributions and also account for its magnitude dependence. As expected, most discrepancies appear
towards the bright end (i.e. $i \lesssim 19$) where number counts are small.

To quantify the agreement between the model and the data, in the context 
of the computation of the LF, we compare the observed number counts with those predicted by our model
 in Figure~\ref{fig:nc_adm_z}, for 12 redshift slices from $z=0.54$ to 
$z=0.65$. The subplots provide the fractional residuals, where the errors in the model are obtained from 
the scatter on the Schechter best-fit parameters (see following sections). The agreement between the 
model and the data is quite good across the entire redshift range, within 
the apparent magnitude range that contains the great majority of the sample. Only at the very bright end, 
where Poisson statistics dominate, do we find some notable discrepancies. 
The average of the RMS of the fractional residuals within the redshift range $0.52 < z < 0.65$ is $\sim3.8\%$
for $i>19$ and $\sim6.7\%$ for $i>18.5$. The $i>19$ limit, for a given redshift slice within our preferred 
redshift range, encompasses between $90\%$ and $97\%$ of the entire sub-sample, depending on the redshift slice considered, from low 
to high redshift. The $i>18.5$ limit encompasses at least $99\%$
of any given subsample within the redshift range considered. In order to illustrate how bright these ranges are, 
the vertical dashed line indicates the apparent magnitude at which the corresponding K-corrected absolute magnitude equals $-24$, i.e. $^{0.55}M_i =  -24$.

Even though the data are noisy at the very bright end ($i\sim18$), there might be a
systematic tendency for the model to under-predict the data. This situation might be due to the presence of 
a small number of unaccounted non-CMASS objects that artificially pass the CMASS selection and only show up where 
legitimate CMASS galaxies 
are scarce. It is also possible that the Schechter function is intrinsically not a good model for
the extremely bright end of the RS LF. We have checked that this discrepancy has 
no effect on our results, as the number densities where 
this effect is noticeable are extremely low and not utilized to extract any conclusion.

\begin{figure}
\begin{center}
\includegraphics[scale=0.57]{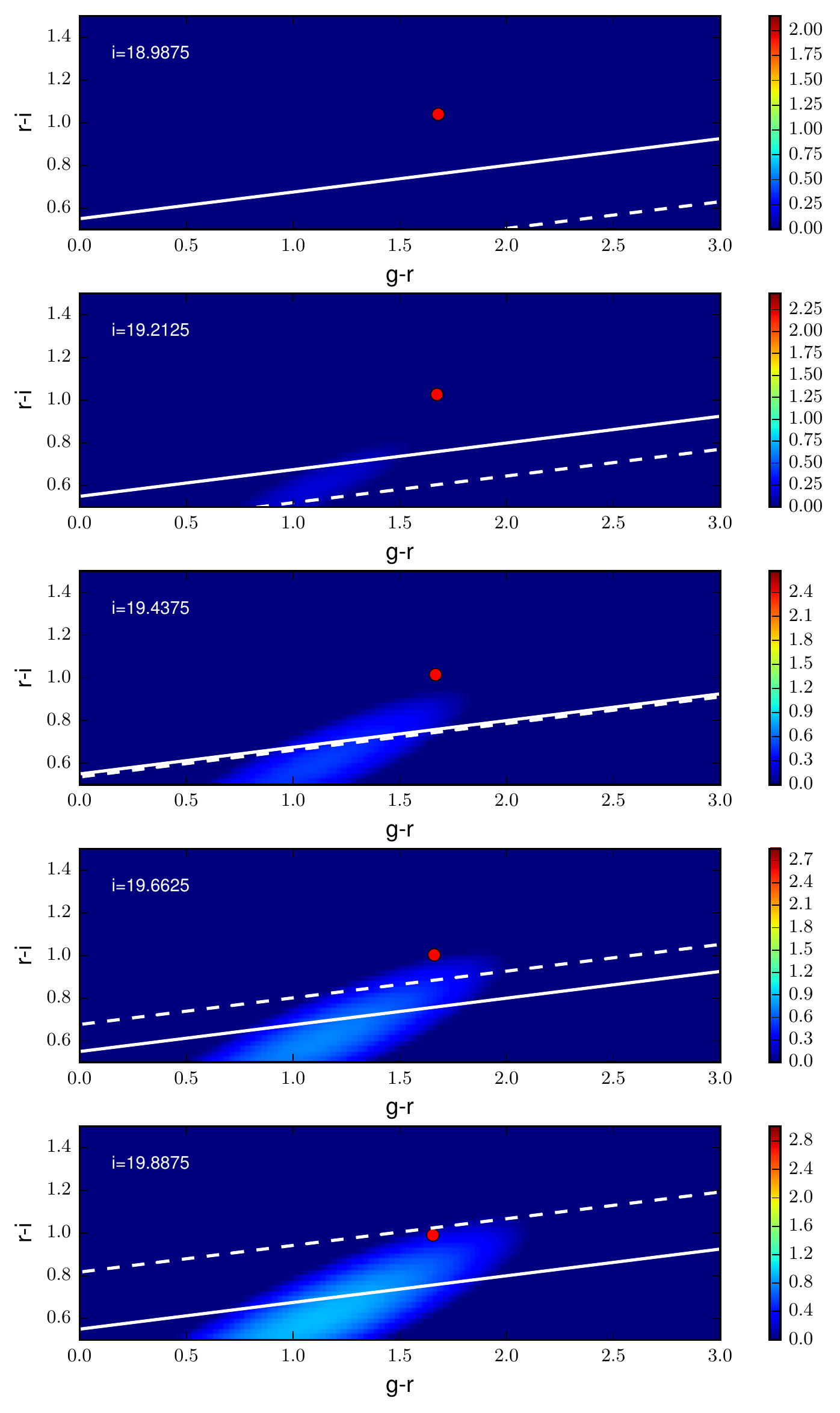}
\caption{Best-fit intrinsic distributions in the colour-colour diagram at $z=0.55$ for 5 i-band apparent magnitude bins 
(in logarithmic scale). The point representing the RS has been exaggerated for the sake of clarity. In 
each panel, the solid and dashed lines represent the $d_\bot$ demarcation $d_\bot > 0.55$, and the sliding cut 
$i < 19.86 + 1.6 \times (d_\bot - 0.8)$, respectively. The RS is, by definition, a point that shifts slightly towards redder colours for fainter magnitudes. The BC, on the other hand, is 
consistent with a diffuse distribution that extends through the red side of the
diagram, where the RS is superimposed upon.}
\label{fig:int_models055}
\end{center}
\end{figure}

\subsection{The intrinsic high-mass RS: an extremely compact distribution in colour-colour-magnitude space}
\label{sec:intrinsic}

In Figure~\ref{fig:int_models055}, we show the best-fit intrinsic distributions in a colour-colour diagram at $z=0.55$
in progressively fainter i-band magnitude bins in logarithmic scale. The RS is, by definition, a point 
that shifts slightly towards redder colours for fainter magnitudes. The BC, on the other hand, is 
consistent with a diffuse distribution that extends through the red side of the
diagram, where the RS is superimposed upon. These two distinct distributions are identified by exclusively 
using the data that we have in hand, including the error model obtained from Stripe 82 multi-epoch data.

\begin{figure}
\begin{center}
\includegraphics[scale=0.4]{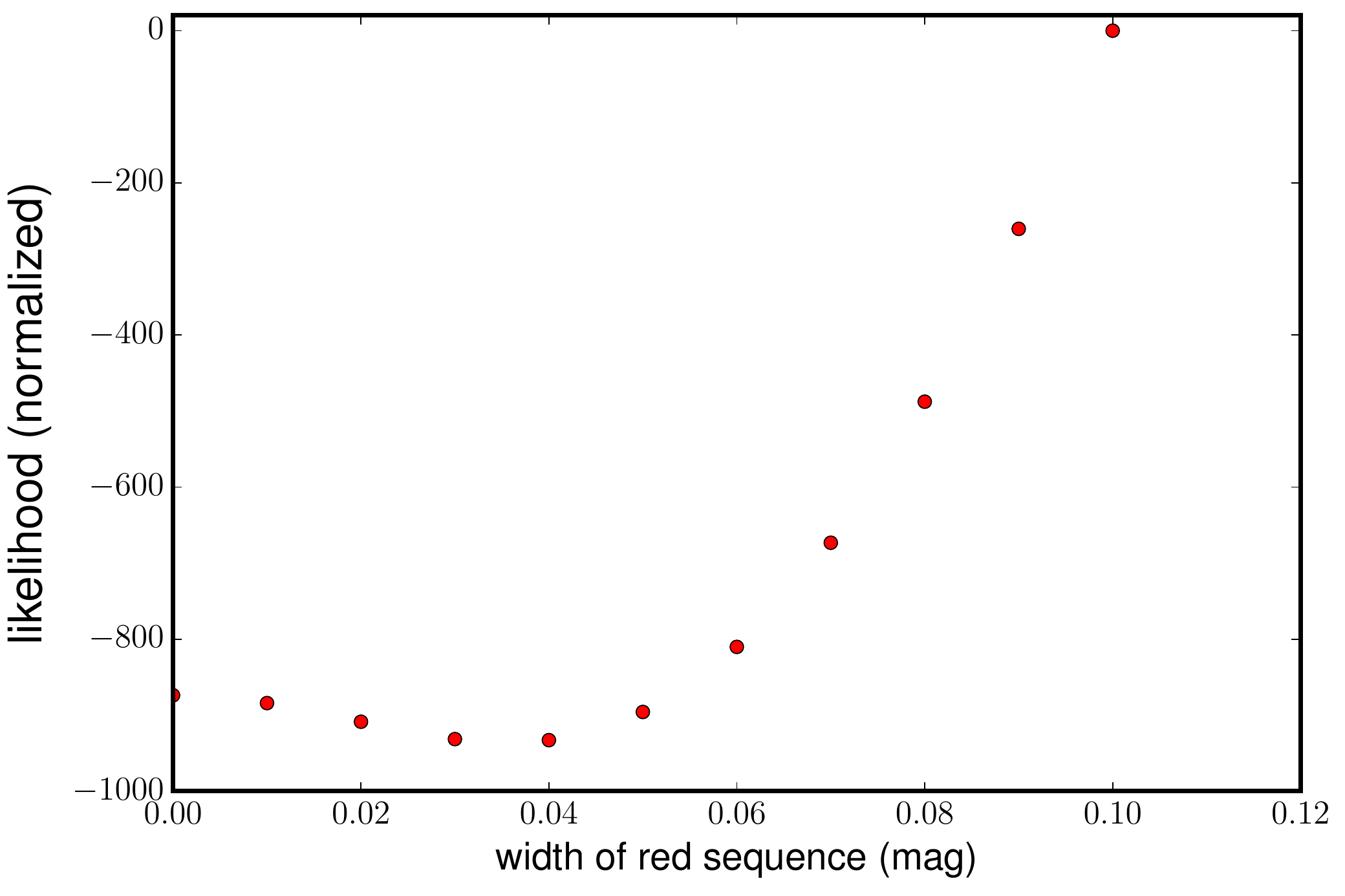}
\caption{Normalized $-{\rm log} \mathscr{L}$ as a function of the width of the RS component 
in the (g-r) vs. (r-i) colour-colour plane at $z=0.55$. }
\label{fig:rs_size}
\end{center}
\end{figure}

The assumption that the high-mass RS is well described by a delta function in the colour-colour plane is motivated 
by our preliminar histogram-deconvolution results. Our analytical deconvolution results clearly confirm
the adequacy of such an assumption. Our modeling gives us robust indications that 
the width of the RS cannot exceed $\sim 0.05$ mag. This result is illustrated in
Figure~\ref{fig:rs_size}, where the likelihood of the model (in the form $-{\rm log} \mathscr{L}$) is shown 
as a function of the width of the RS component for the redshift slice centered at $z=0.55$. The 
minimum likelihood is reached somewhere between $0.03$ and $0.04$ mag, but below 
$0.05$ mag the likelihood function becomes fairly flat. For a width $>0.05$ mag the likelihood
worsens rapidly. The fact that the likelihood function is so flat below $0.05$ mag, 
in combination with the limited resolution that we have (given the large photometric
errors, especially along the g-r colour axis), makes our point-like assumption well justified. 
We have also checked that assuming a width of $0.035$ mag would have a negligible 
effect on the results and conclusions of this work. 

\begin{figure}
\begin{center}
\includegraphics[scale=0.32,angle=90]{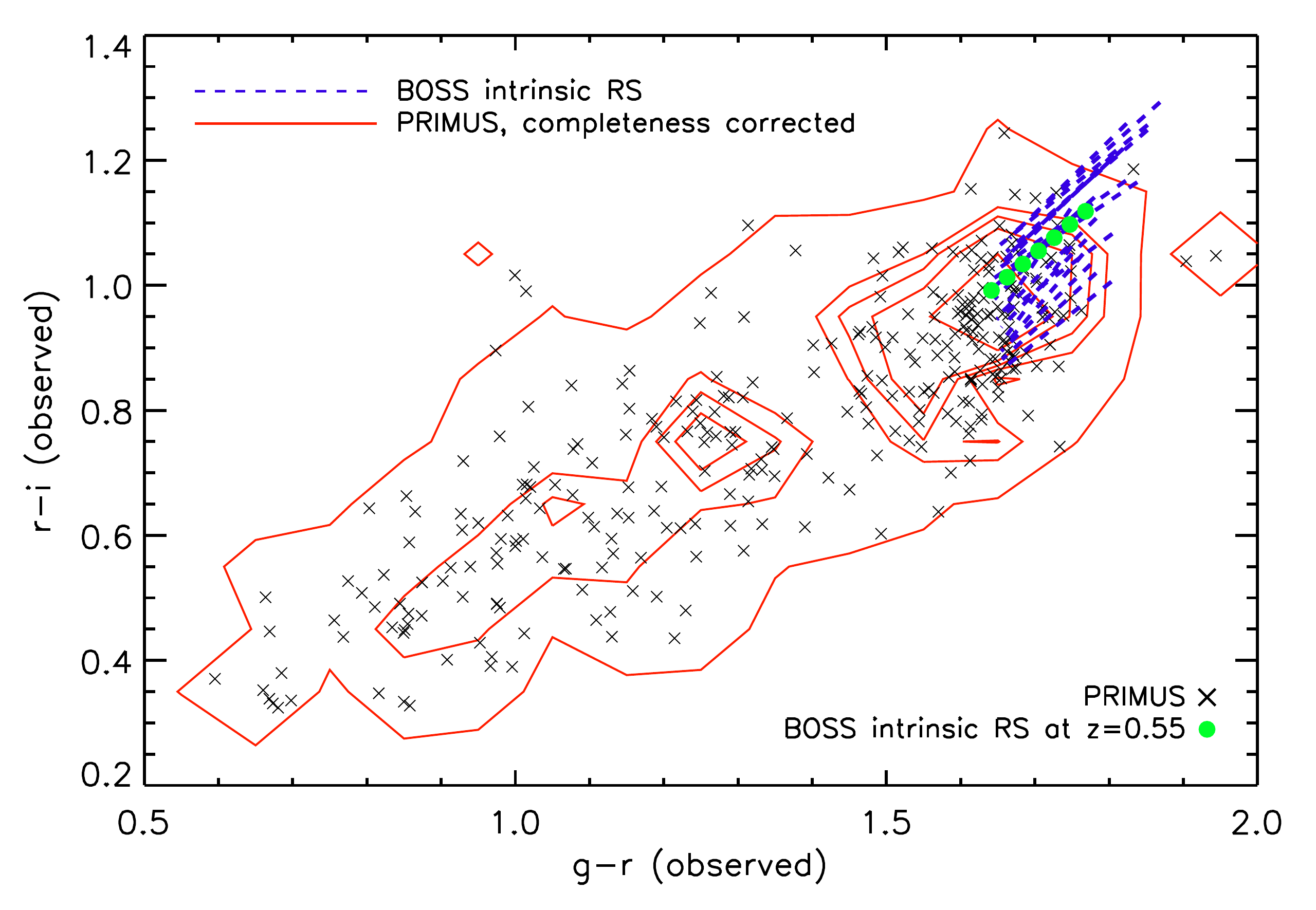}
\caption{The g-r vs. r-i colour-colour diagram for PRIMUS data within the redshift
range $0.5\leq z \leq 0.6$ and magnitude range $17.5 \leq m_i \leq 20$, as compared with the
intrinsic RS location as inferred from this work. Crosses show each of the objects 
contained in the PRIMUS dataset according to the aforementioned selection whereas
the contours include a completeness correction (provided by the PRIMUS team). The blue dashed
lines show the location of the intrinsic RS as inferred from BOSS within the same magnitude 
range for each redshift slice between $z=0.5$ and $z=0.6$. The green dots 
represent the $z=0.55$ results.}
\label{fig:primus}
\end{center}
\end{figure}

We have compared our results for the location of the intrinsic RS colour-colour
distribution with publicly available PRIMUS data, to the extend that the data allow. This 
comparison is illustrated in a g-r vs. r-i colour-colour diagram in Figure~\ref{fig:primus}. 
The position of the RS as inferred from both surveys is in good agreement, even 
though the CMASS RS is slightly redder. Note that this 
is a qualitative comparison, as in the magnitude range covered by the CMASS sample the 
PRIMUS data are very scarce. Also, the PRIMUS distribution is affected  
by photometric errors. Importantly, Figure~\ref{fig:primus} illustrates that the scatter found in 
the PRIMUS data can be understood in terms of the colour-magnitude
relation and the progressive reddening produced by redshift. 

Independently of the priors that we choose, our model requires an intrinsic BC
component that extends through the red side of the colour-colour diagram, in order to achieve 
an acceptable agreement with the number counts. This result implies that there are BC galaxies that 
are intrinsically redder than the RS itself. This
should not be interpreted in light of any previous arbitrary separation between red and blue
galaxies, based on colour cuts or stellar population properties. Again, the BC distribution
must be considered a background component, that includes every type of object which is not contained in the 
extremely-narrow RS component. By looking at available BOSS spectra, we have checked that the red side of the BC distribution
is spectroscopically heterogenous, to the extent that BOSS S/N permits. 
The existence of spirals with unusually red colours has been thoroughly documented in the 
literature (e.g., \citealt{Masters2010, Masters2010b},  \citealt{Wolf2009}, \citealt{Bundy2010}). 
Also, the definition of our BC as a background distribution containing all objects that do not 
belong to the extremely concentrated RS, is consistent with the presence of dusty early-type galaxies. 
In following sections, the expected fraction of BC objects within the entire sample, along with 
the fraction of BC objects on the red side of the colour-colour plane will be provided.

\subsection{Completeness in the CMASS sample}
\label{sec:completeness}

Our deconvolution method allows us to estimate completeness as a function of intrinsic magnitude, colour and redshift. 
Completeness is defined here as {\it{the fraction of objects from the intrinsic distribution (in intrinsic space)
within a given range of magnitude, colour and redshift that is expected to pass the CMASS selection}}. 

\begin{figure}
\begin{center}
\includegraphics[scale=0.40]{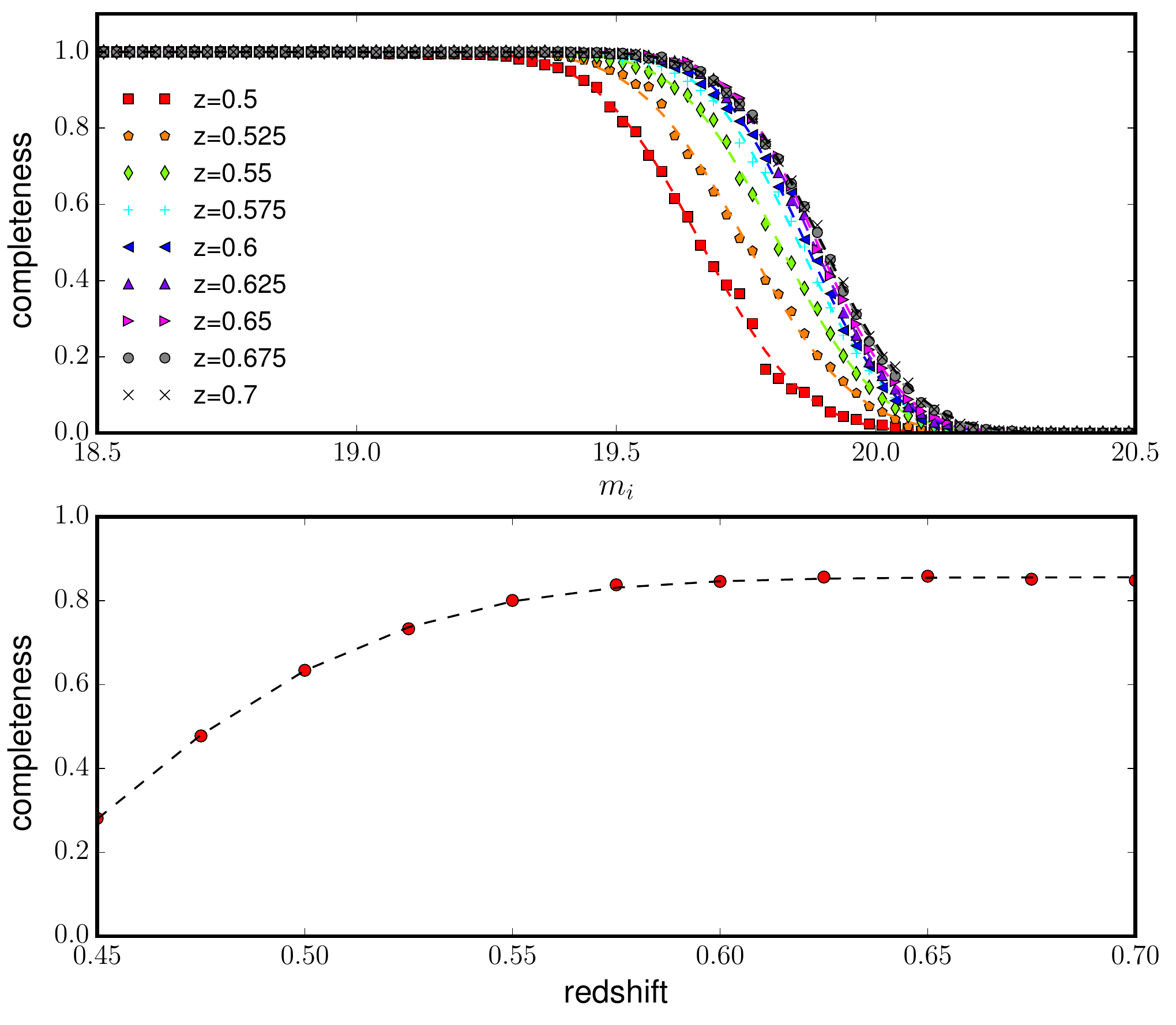}
\caption{Completeness in the CMASS sample. Upper panel: RS completeness as a function of i-band magnitude in the 
CMASS sample for nine redshift slices within the redshift range $0.5<z<0.7$. Lower panel: Overall RS completeness 
as a function of redshift, integrating over the colour and apparent magnitude limits of the 
CMASS selection. As expected, the sample is complete for the RS
at the bright end ($i\lesssim19.4$) across the entire redshift range. Between $i\simeq19.4$ and $i\simeq19.7$, completeness
declines quite abruptly, this effect occurring at progressively fainter magnitudes for
higher redshifts. The overall RS completeness increases from a very 
low value at low redshift to $\sim 0.8$ at $z=0.55$, reaching a plateau around $\sim0.82$ at $z\gtrsim0.52$. Analytical 
forms for completeness as a function of magnitude and redshift (dashed lines) are given by Equation~\ref{eq:comp_mag} and 
Equation~\ref{eq:comp_z} and best-fit parameters are listed in Table~\ref{table:comp_fits}.}
\label{fig:completeness}
\end{center}
\end{figure}

In the upper panel of Figure~\ref{fig:completeness} we show completeness as a function of magnitude for the RS distribution in 
nine redshift slices from $z=0.50$ to $z=0.70$. As expected, the sample is complete for the RS
at the bright end ($i\lesssim19.4$) across the entire redshift range. Between $i\simeq19.4$ and $i\simeq19.7$, completeness
declines abruptly, this effect occurring at progressively fainter magnitudes for
higher redshifts. The shape of the completeness function is determined by the $d_\bot$ cut and the 
sliding colour-magnitude cut being applied to a RS distribution that is progressively redder 
at higher redshifts. The bottom panel of Figure~\ref{fig:completeness} provides the 
integrated RS completeness of actual CMASS sample relative to what it would be, given noise-free 
photometry as a function of redshift. Here, we have integrated over the colour and 
magnitude limits of the CMASS selection. The RS completeness increases from a small
value at low redshift to $\sim 0.8$ at $z=0.55$, reaching a plateau of $\sim0.82$ at $z\gtrsim0.52$. 
Unfortunately, we can only confidently report completeness for the RS, due to the extreme incompleteness on the BC.

The RS completeness as a function of apparent magnitude for a given redshift slice $z$ is well fit by the following 
function (dashed lines in Figure~\ref{fig:completeness}):

\begin{equation} 
C(m_i) = \frac{1}{2} \mathrm{erfc} \left [\frac{m_i - c_m^0 }{c_m^1} \right]
\label{eq:comp_mag}
\end{equation}

where $\mathrm{erfc}$ is the complementary error function. Similarly, the integrated RS completeness as a function of redshift is well fit by the following expression:

\begin{equation}
C(z) = \frac{1}{2} c_z^0 \mathrm{erf} \left [\frac{m_i - c_z^1 }{c_z^2} \right]
\label{eq:comp_z}
\end{equation}

where $\mathrm{erf}$  is the error function. Table~\ref{table:comp_fits} lists best-fit coefficients for completeness as a function of magnitude 
(for 9 different redshift slices) and integrated completeness as a function of redshift.

\begin{table}

\begin{center}
\small{
\begin{tabular}{cccc}
\hline
\hline
	         Redshift&   $c_m^0$&   $c_m^1$     \\
\hline
0.450 &      $19.133$ &      $ 0.800$\\
\hline
0.475 &      $19.533 $ &     $0.235$\\
\hline
0.500 &      $19.662$ &      $0.224$\\
\hline
0.525 &      $19.750$ &      $0.233$ \\
\hline
0.550 &      $19.813$ &      $0.213$\\
\hline
0.575 &      $19.851$ &      $0.200$\\
\hline
0.600 &      $19.864$ &      $0.189$\\
\hline
0.625 &      $19.880$ &      $0.183$\\
\hline
0.650 &      $19.887$ &      $0.186$\\
\hline
0.675 &      $19.893$ &      $0.199$\\
\hline
0.700 &      $19.895$ &      $0.204$\\
\hline
\hline
		 $c_m^0$&   $c_z^1$&   $c_z^2$   \\
\hline
$1.712$&  $0.420$&  $0.100$	\\			
\hline
\end{tabular}}
\end{center}
\caption{Best-fit parameters for red sequence completeness as a function of i-band apparent 
magnitude for 9 different redshift slices, according to the analytical form of 
Equation~\ref{eq:comp_mag}. Also listed are the best-fit parameters 
for the integrated red sequence completeness as a function of redshift, according to Equation~\ref{eq:comp_z}. In both cases,
results below $z=0.5$ must be considered pure extrapolation of the higher-redshift results, as 
completeness is too low to allow for a reliable deconvolution of the intrinsic populations.}
\label{table:comp_fits}
\end{table}

As previously pointed out, it would be risky to report completeness/LF for the BC, given 
the insufficient constraints in the CMASS data. However, we can 
provide an  estimation of the fraction of BC objects in the sample, i.e., the $f_{blue}$ parameter, and
the contamination from intrinsically BC objects on the red side of the colour-colour diagram.
We emphasize here the difference between {\it{the red side}}, which is commonly defined
by a simple colour demarcation (e.g., $g-i = 2.35$) in observed space, and the RS, which is a distinct 
galaxy population, modeled here as a delta function in the colour-colour plane, and 
intrinsically separated from the BC component. 

\begin{figure}
\begin{center}
\includegraphics[scale=0.5]{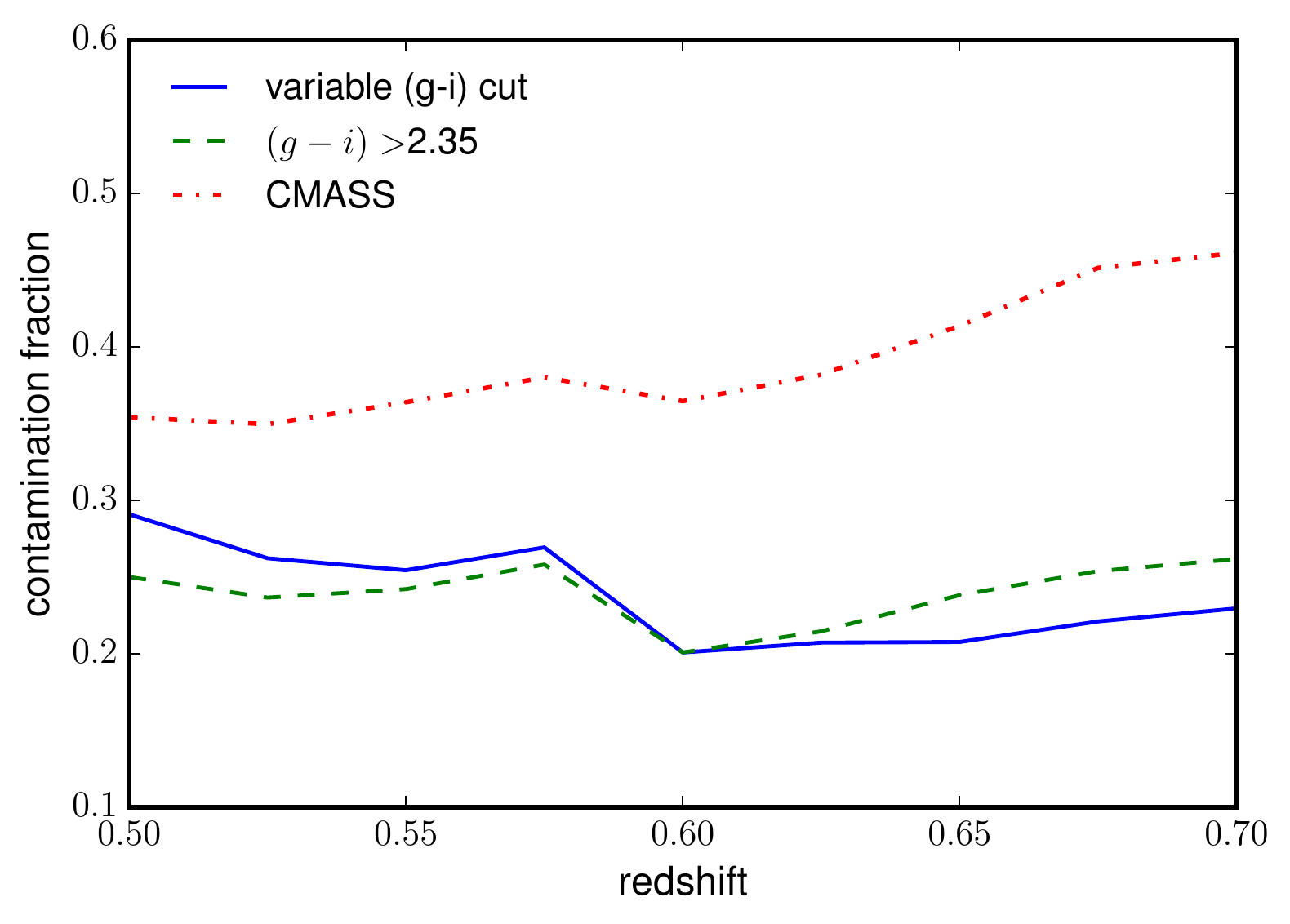}
\caption{The intrinsic fraction of BC objects contaminating the red side of the colour-colour diagram of the total 
number of red objects. The red side is defined here in two different ways: using a fixed redshift-independent  
g-i demarcation at $g-i = 2.2$, $g-i = 2.3$ and $g-i = 2.4$, respectively, and using a redshift-dependent 
demarcation that is obtained empirically from the observed colour-colour distributions. Also shown 
in the plot is the total fraction of BC objects in the CMASS.}
\label{fig:contamination}
\end{center}
\end{figure}

Figure~\ref{fig:contamination} shows contamination as a function of redshift assuming two g-i colour demarcations. Firstly,
we use a fixed g-i cut at $g-i = 2.35$, which has been previously used to separate 
red and blue galaxies in the CMASS sample in observed space \citep{Maraston2013}. Under such condition, the contamination hovers around
$\sim25\%$ within the redshift range considered. In order to account for the effect of redshift on the colour distributions, we
also assume a redshift-dependent demarcation that is obtained empirically from the observed colour-colour distributions themselves. 
This approach decreases the contamination from $\sim 0.30$ at $z=0.50$ to $\sim 0.22$ at $z=0.70$. Interestingly, 
even though the completeness of the BC increases significantly  
towards higher redshifts (i.e., we have more BC objects at high redshift), the fraction 
of BC objects that contaminate the red side remains essentially constant. Figure~\ref{fig:contamination} also displays
the fraction of BC objects in the sample (the $f_{blue}$ parameter). This fraction does experience a considerable increase 
as a function of redshift, from $\sim36\%$ at $z=0.50$ to $\sim46\%$ at $z=0.70$. The total fraction of intrinsic BC objects (not just objects with blue colours) in the 
CMASS sample is $37\%$.

\subsection{The high-mass end of the Red Sequence Luminosity Function at $z\sim0.55$}
\label{sec:lf}

Our analytical deconvolution method provides, by construction, a straightforward approach
to determine the LF for both the RS and the BC, although (due to BC incompleteness) only the
first one is reported here. We have checked that our main conclusions in terms of colour bimodality and RS completeness 
are not significantly affected by systematic error as quantified by the variation in the $\beta$ parameter
described in Section~\ref{sec:uncertainty}. Likewise, our results for the normalization of the LF are robust.
Our results for the redshift-dependent variation of the LF are more uncertain given
the level of systematic error in our modeling.

As indicated above, the BOSS selection is designed to peak at $z\sim0.55$. When we move away 
from this redshift we encounter several selection effects. At higher redshift, 
the {\it{effective}} magnitude range becomes progressively narrower, with 
number counts being pushed to the faint limit of the survey. Also, the fraction of BC objects in the sample becomes 
significant. At lower redshift, the BC is not accesible at all,
but completeness in the RS is too low due to the colour cut 
to guarantee an efficient deconvolution. The coupling between these redshift-dependent selection
effects, the relatively narrow overall redshift range, and the residual systematics of our modeling
limits the effectiveness of our analysis for the study of RS evolution internal to the CMASS sample.
A better measurement of RS evolution would require
a more accurate determination of the error model in follow-up work, and/or 
the incorporation of higher/lower redshift results into our analysis framework.

\subsubsection{LF computation and the Schechter parametrization}
\label{sec:schech}

The RS LF is modeled as a Schechter function, which is the functional 
form that we have found to describe better the number counts at the bright end. However, the 
entire CMASS sample is considerably brighter 
than the best-fit $m_*$ (and consequently, $M_*$). For this reason, the Schechter parametrization 
that we derive for the RS LF must be considered only an analytic functional form; making physical interpretations
using the Schechter parameters alone can be misleading.

\begin{figure}
\begin{center}
\includegraphics[scale=0.39]{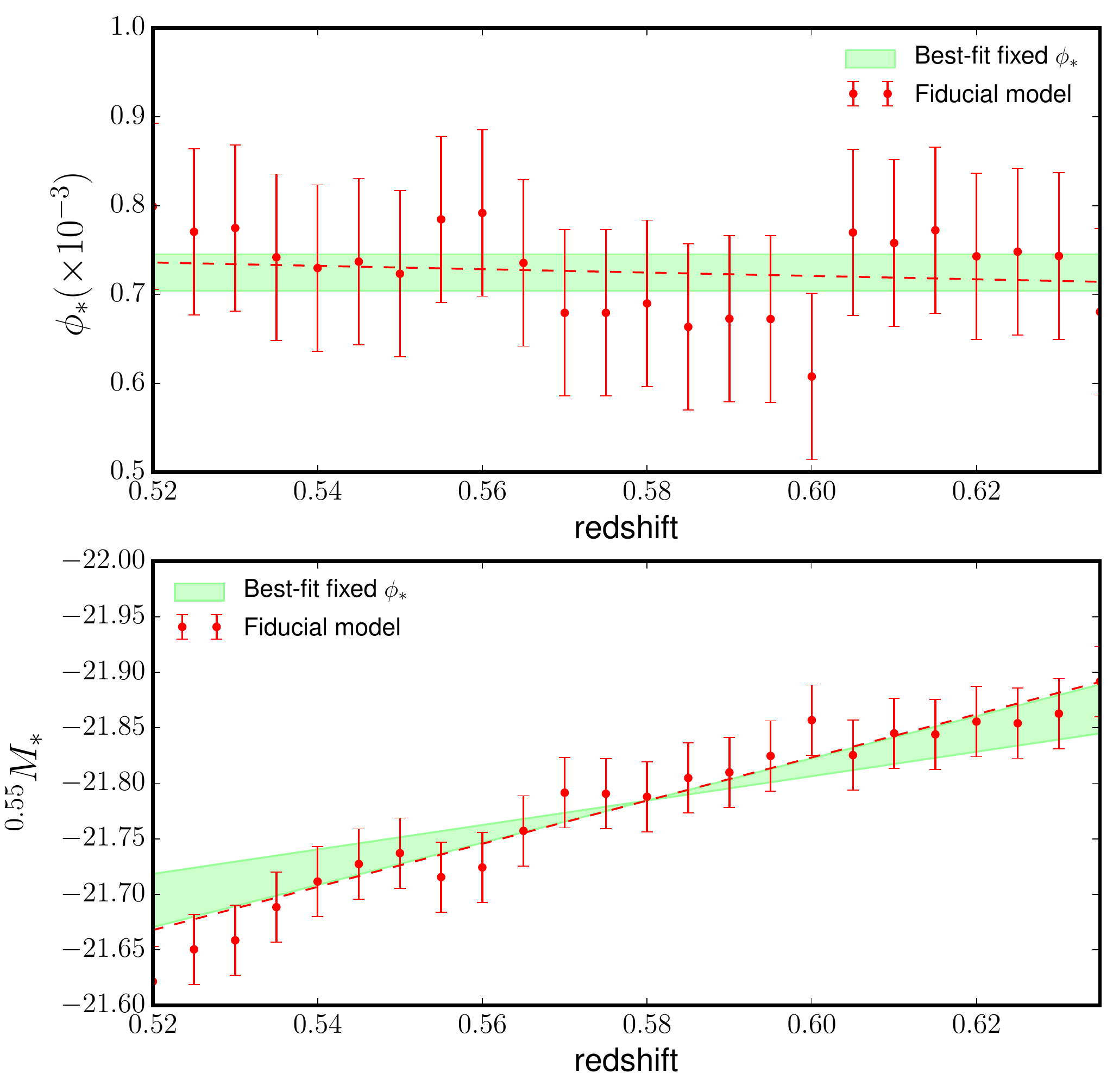}
\caption{Redshift evolution of the computed Schechter parameters $\phi_*$ and 
$^{0.55}M_*$ for the bright end of the RS LF. The red dots represent the fiducial model, and the 
red dashed line, a linear fit to the fiducial model within the redshift range $0.525<z<0.63$. The error bars  
represent the statistical error of the optimization and the uncertainty in the modeling. 
The solid line and shaded regions show the best-fit constant-$\phi_*$ model (purely-passive) model and 
its uncertainty range, respectively.}
\label{fig:sch_param1}
\end{center}
\end{figure} 

We have assumed a fixed value for the faint-end slope, $\alpha = -1$. This fixed choice for $\alpha$ is motivated by the insufficient constraints 
on the faint-end slope that we can extract from the CMASS sample. The $\phi_*$ Schechter 
parameter is estimated for the RS component and for each redshift slice in the way  
described in Section~\ref{sec:computation}. Here, the maximum volume is 
assumed to be constant within the corresponding redshift slice. Neglecting 
any $1/V_{max}$ effects is a reasonable approximation considering the width of our redshift slices and the
 apparent magnitude range of the CMASS selection. The 
typical distance-modulus variation within our redshift slices is $\sim 0.05$ mag, which is of the 
order (if not smaller) of the typical i-band magnitude error.

The absolute-magnitude $M_*$ parameter is directly derived by transforming the best-fit (apparent magnitude) $m_*$ parameter, 
using the standard equation. For the K-correction we opt to blueshift the rest-frame redshift to $z=0.55$.
The motivation for this choice is two-fold. First, the $^{0.55}i$ passband is covered within the entire 
CMASS redshift range. Second, K-correcting to a redshift that is close to the mean redshift of the sample
reduces the correction itself and, therefore, the uncertainties associated with it. We also 
choose not to apply any evolution correction ( that accounts for the intrinsic luminosity evolution of each galaxy 
from $z$ to the chosen rest-frame redshift). The size of the CMASS sample allows us to 
split the sample in very narrow redshift bins, where evolutionary effects are negligible. 

Average K-corrections to $z_0=0.55$ are derived, for each bin in the colour-colour plane, 
using the extensive grid of physically plausible FSPS models described in Section~\ref{sec:expected_distribution}.
The fact that the RS intrinsic distribution is modeled as a delta function on the
colour-colour plane helps simplify the computation here. Moreover, the computed 
displacement of the RS centroid within the CMASS magnitude range is rather small (i.e., $<0.1$),
so we can simply take the average of the K-correction within this magnitude range. 
We have checked that accounting for the colour-magnitude relation for the RS 
when estimating K-corrections does not change our results for the LF in any significant way.

\begin{table}
\begin{center}
\small{
\begin{tabular}{cccc}
\hline
\hline
	         Redshift&   $\phi_* \times 10^{-3}$ Mpc$^{-3}$ mag$^{-1}$&   $^{0.55}M_*$     \\
\hline
0.525 &      $ 0.77 \pm 0.0930 $ &      $ -21.6504 \pm 0.047 $\\
\hline
0.53 &      $ 0.7749 \pm 0.0938 $ &      $ -21.6586 \pm 0.0483 $\\
\hline
0.535 &      $ 0.7420 \pm 0.0910 $ &      $ -21.6885 \pm 0.0490 $\\
\hline
0.54 &      $ 0.7298 \pm 0.0903 $ &      $ -21.7115 \pm 0.0489 $\\
\hline
0.545 &      $ 0.7371 \pm 0.0878 $ &      $ -21.7272 \pm 0.0483 $\\
\hline
0.55 &      $ 0.7234 \pm 0.0873 $ &      $ -21.7371 \pm 0.0491 $\\
\hline
0.555 &      $ 0.7846 \pm 0.0991 $ &      $ -21.7154 \pm 0.0494 $\\
\hline
0.56 &      $ 0.791 \pm 0.101 $ &      $ -21.7242 \pm 0.0501 $\\
\hline
0.565 &      $ 0.735 \pm 0.0956 $ &      $ -21.7572 \pm 0.0511 $\\
\hline
0.57 &      $ 0.679 \pm 0.0920 $ &      $ -21.7915 \pm 0.0522 $\\
\hline
0.575 &      $ 0.6 \pm 0.0973 $ &      $ -21.7906 \pm 0.0539 $\\
\hline
0.58 &      $ 0.690 \pm 0.110 $ &      $ -21.7878 \pm 0.0586 $\\
\hline
0.585 &      $ 0.6636 \pm 0.105 $ &      $ -21.8047 \pm 0.0570 $\\
\hline
0.59 &      $ 0.6728 \pm 0.107 $ &      $ -21.8098 \pm 0.0588 $\\
\hline
0.595 &      $ 0.6724 \pm 0.105 $ &      $ -21.8246 \pm 0.0604 $\\
\hline
0.6 &      $ 0.6077 \pm 0.106 $ &      $ -21.8568 \pm 0.061 $\\
\hline
0.605 &      $ 0.7698 \pm 0.102 $ &      $ -21.8253 \pm 0.0630 $\\
\hline
0.61 &      $ 0.7580 \pm 0.0986 $ &      $ -21.8450 \pm 0.0524 $\\
\hline
0.615 &      $ 0.7724 \pm 0.12 $ &      $ -21.8440 \pm 0.0661 $\\
\hline
0.62 &      $ 0.7430 \pm 0.125 $ &      $ -21.8555 \pm 0.066 $\\
\hline
0.625 &      $ 0.7482 \pm 0.137 $ &      $ -21.8540 \pm 0.0687 $\\
\hline
0.63 &      $ 0.7433 \pm 0.141 $ &      $ -21.8627 \pm 0.0682 $\\
\hline
\end{tabular}}
\end{center}
\caption{Best-fit Schechter parameters for the i-band RS LF at multiple redshift slices from $z=0.525$ to
$z=0.63$. The LF has been k-corrected to $z=0.55$. We have assumed $\alpha=-1$. }
\label{table:schechter_fits}
\end{table}

Figure~\ref{fig:sch_param1} shows the redshift evolution of best-fit Schechter 
parameters $\Phi_*$ and $^{0.55}M_*$, within the redshift range 
$0.525<z<0.63$. The best-fit linear relations are:

\begin{eqnarray} \displaystyle
\Phi_* = [(-0.189\pm0.372)~z + (0.834\pm0.216)] \times \\ 
\times10^{-3}~{\rm{Mpc}}^{-3}~{\rm{mag}}^{-1}\nonumber \\ 
^{0.55}M_* = (-1.943\pm0.228)~z  + (- 20.658\pm0.132)
\label{eq:linear_fits2}
\end{eqnarray}

The error bars shown in Figure~\ref{fig:sch_param1} take into account not only the statistical errors of the 
optimization procedure, but also the uncertainty in the modeling, derived from the variation 
of the best-fit $\beta$ parameter (see Section~\ref{sec:uncertainty}). We have opted to use a restricted redshift range in 
order to minimize the effect of systematics on the computation of the RS LF at high and low redshift. Table 
\ref{table:schechter_fits} lists the Schechter parameters within this redshift range.

Figure~\ref{fig:sch_param1} shows a very mild evolution 
in $\Phi_*$ within the narrow redshift 
range considered, with a slope for the best-fit linear relation of $-0.189\pm0.372$. In a shaded region, we provide 
our estimate for the best-fit purely-passive model (constant $\Phi_*$), i.e. 
$\Phi_* = (7.248 \pm 0.204)\times10^{-4}$ Mpc$^{-3}$ mag$^{-1}$. 
The $^{0.55}M_*$ - z relation corresponding to the best-fit fixed-$\Phi_*$ model is also 
represented by a shaded region in the bottom panel of Figure~\ref{fig:sch_param1}. 
We estimate that a passive model with a fading rate within the range $\Delta ^{0.55}M_* /\Delta z = [1.1-1.9]$ is consistent with 
both the data and the uncertainties in the modeling.

\begin{figure*}
\begin{center}
\includegraphics[scale=0.6]{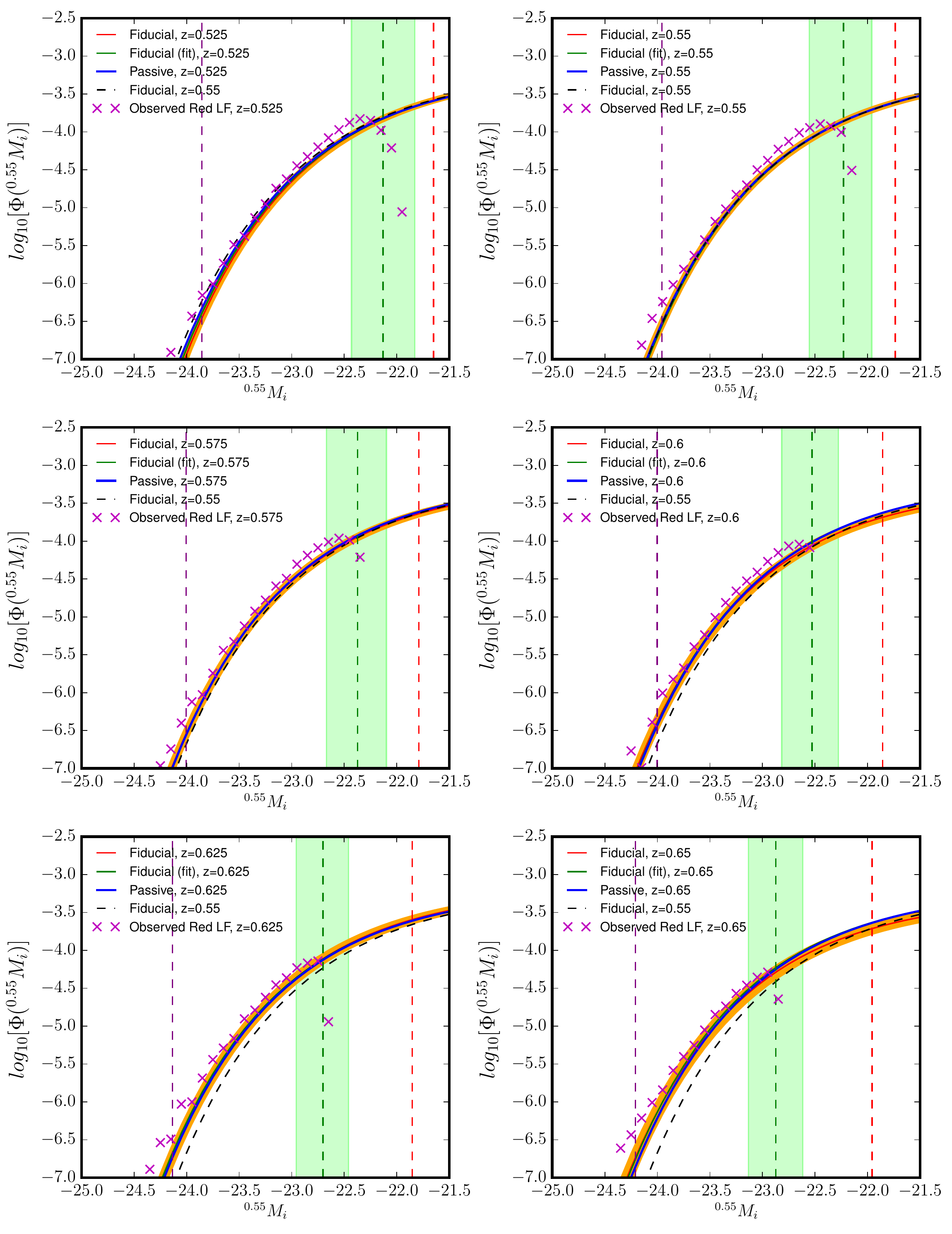}
\caption{The bright end of the $^{0.55}i$-band RS LF in six redshift slices between $z=0.525$ and
$z=0.65$. Each panel shows the RS LF computed with our fiducial model (solid red line), the RS LF obtained with the linear fits to the Schechter parameters 
(solid green line) and our best-fit passive-evolving model (solid blue line). In addition,
we show for reference the fiducial model at $z=0.55$ (dashed black line). Vertical 
dashed lines represent, from left to right: a bright-end reference magnitude corresponding to 
the value below which the total number of CMASS objects is 25 (purple line); the $0.5$-completeness limit of the sample (green line, including 
in a shaded region the $0.02$ and $0.98$ completeness limits) and $^{0.55}M_*$ (red). The crosses 
represent the RS LF previous to any deconvolution (see text). Orange shaded 
regions show the uncertainty in our RS LF estimate, taking into account both the systematic and the statistical error.}
\label{fig:lf_all}
\end{center}
\end{figure*}

Due to the fact that the Schechter parameters are covariant, relatively large 
variations in the Schechter parameters translate into small variations in the 
shape of the LF. Figure~\ref{fig:lf_all} presents the RS LF in 6 redshift bins within 
the redshift interval $0.525<z<0.65$. In each panel, we show the RS LF computed 
with our fiducial model at the corresponding redshift, the RS LF obtained with the linear fits to the Schechter parameters of
 the fiducial model at the corresponding redshift, our best-fit passive-evolving model with $\Delta ^{0.55}M_* /\Delta z = 1.5$ (the
center value of our high-confidence interval), 
the fiducial model at $z=0.55$ and an ``observed" RS LF previous to any deconvolution, assuming a fixed colour demarcation of $g-i = 2.35$
to separate the red and blue populations. Figure~\ref{fig:lf_all} illustrates the extreme luminosity of the sample, with 
$50\%$ completeness limits between $-22.1$ and $-22.7$, i.e., $0.6-1$ mag brighter 
than $^{0.55}M_*$ within our high-confidence redshift interval. Importantly, 
Figure~\ref{fig:lf_all} shows that the evolution of the massive-end RS LF is 
consistent with our best-fit passive-evolution model, with $\Phi_* = (7.248\pm0.204)\times10^{-4}$ Mpc$^{-3}$ mag$^{-1}$ and 
a fading rate of $\Delta ^{0.55}M_* /z \simeq 1.5$ per unit redshift (recall that values 
between $\sim1.1$ and $1.9$ are possible within the uncertainties of the modeling). Deviations
from passive evolution are only noticeable at the very brightest end, and at the edges of the redshift range, and are 
always within the estimated uncertainties in the computation.

A constant-$\Phi_*$ result at the high-mass end contrasts with the measured evolution of $\Phi_*$ at intermediate 
masses. Results obtained from samples that directly explore the $M_*$ and $\Phi_*$ portion of the LF 
indicate that $\Phi_*$ for the intermediate-mass red galaxy population increases by at least a factor 2 from $z=1$ (see, e.g., 
\citealt{Faber2007}).

\subsubsection{Comparison with previous LF evolution results at $z\sim0.5$}
\label{sec:comparison}

\begin{figure}
\begin{center}
\includegraphics[scale=0.27]{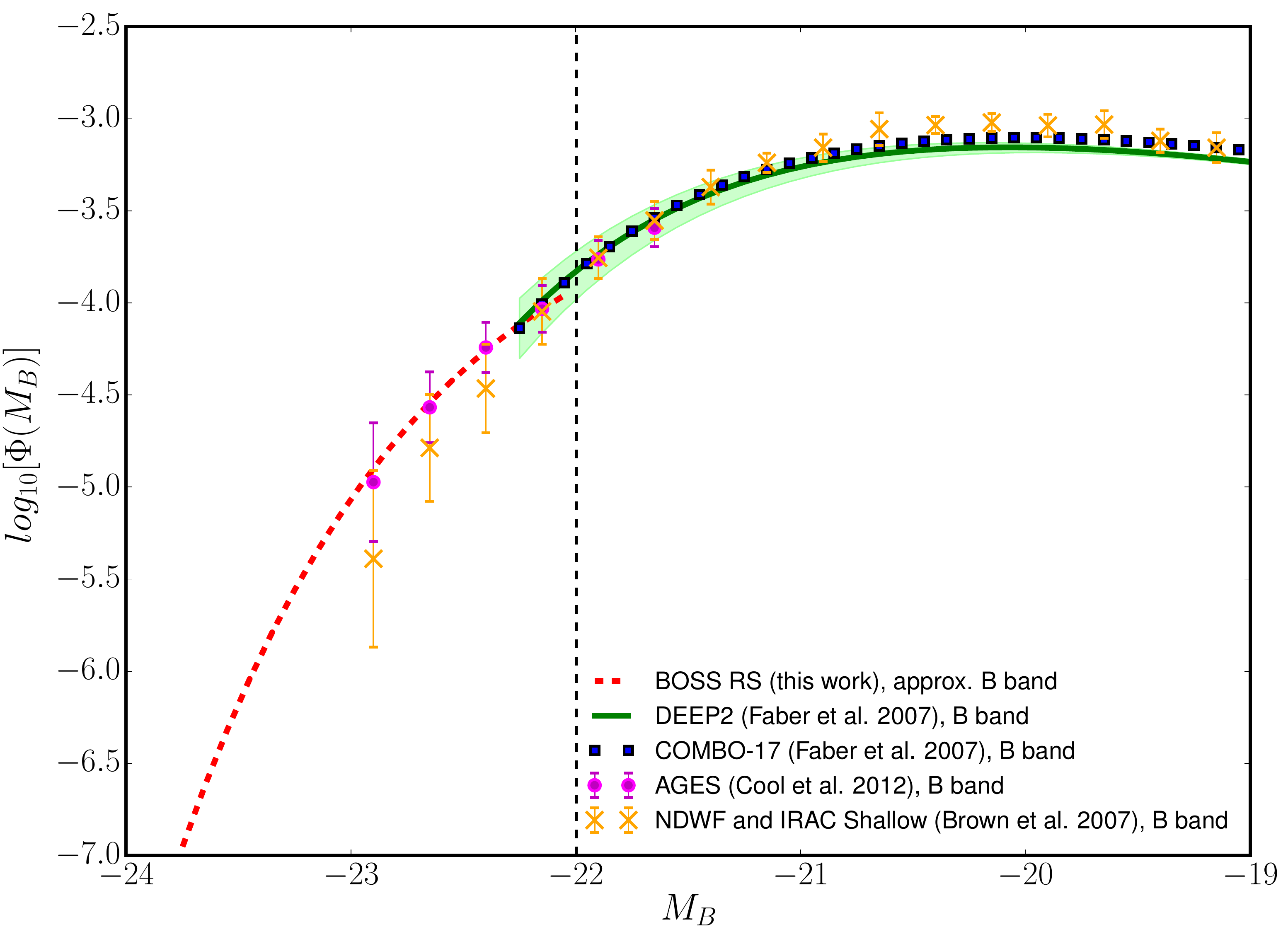}
\caption{The CMASS RS LF at $z=0.5$ in the B band (approximately) as compared with 
previously reported LFs for red galaxies. The $^{0.55}i$ CMASS RS LF at $z=0.5$ is 
transformed into the rest-frame (Johnson) B-band LF, by taking advantage of the fact that the i-band filter at $z=0.55$ approximates the
SDSS g band. The g-band magnitude is transformed into the B-band magnitude following Equation~\ref{eq:B}. The following 
red LFs are used in this comparison: the DEEP2 Red 
LF \citep{Faber2007}, the Red LF in the Bo\"otes field from \citep{Brown2007} using the NDWF and {\it{Spitzer}} IRAC Shallow surveys, the Red LF in the Bo\"otes field from \citep{Cool2012} using AGES and
the COMBO17 Red LF \citep{Faber2007}. The vertical line shows the approximate magnitude where completeness in the CMASS sample drops to $50\%$. 
For the DEEP2 and the COMBO17 LFs, the Schechter fits to the data points are plotted, and $M_B = 22.15$ 
is used as the bright limit of the measurements (see \citealt{Faber2007}, Figure 6). Our RS LF is a continuation of the 
low-mass RS LFs, within the errors reported by the aforementioned works.}
\label{fig:lf_comparisons}
\end{center}
\end{figure}

Due to footprint limitations, previous surveys are severely affected by 
small-number statistics in the absolute magnitude ranges probed by BOSS. Here we show how our high-mass estimate compares with published lower-mass 
RS LFs to the extent that the data allow. This comparison is presented in Figure~\ref{fig:lf_comparisons} for a representative 
set of previous results. As most LFs in the literature are presented in the B band, the $^{0.55}i$ CMASS RS LF at $z=0.5$ is 
transformed into the rest-frame (Johnson) B-band LF, by taking advantage of the fact that the i-band filter at $z=0.55$ approximates the
SDSS g band. We apply the following transformation from g-band magnitudes to B-band magnitudes:

\begin{equation}
B = g + 0.115 + 0.370 \times (g-r)
\label{eq:B}
\end{equation}

\noindent which is taken from \cite{Bell2004} and \cite{Faber2007}. Note that the redshift slice $z=0.50$ is slightly outside our high-confidence
range, so Schechter parameters for the CMASS RS LF are obtained by using the fits 
of Equation~\ref{eq:linear_fits2}. Note that the comparison  
shown in Figure~\ref{fig:lf_comparisons} is only approximative, as it involves several assumptions that can shift 
our LF slightly along the x axis. This said, our RS LF is a continuation of the 
low-mass RS LFs, within the errors reported by the aforementioned works. At the bright end, most surveys, as mentioned before, are very incomplete. Our results are in good agreement with those from \cite{Cool2012}, down to $M_B \sim -23$. 

Figure~\ref{fig:lf_comparisons} confirms that our methodology for inferring the LF of RS galaxies from BOSS
data, including the treatment of photometric errors, selection cuts, and red-blue
deconvolution, is sound and yields measurements that define a consistent picture 
when viewed alongside previous measurements. It also shows the unique capability of 
BOSS for studying the most massive tail of the massive galaxy population.

\subsubsection{Quantifying the evolution of LRGs: Is passive evolution a good approximation?}
\label{sec:passive}

Although our method has the potential to provide a definitive answer to this question, due to the 
narrow redshift range and the sensitivity of the Schechter parameters to uncertainties 
in the error model, we can only state that our results are {\it{consistent}} with passive evolution. In this section 
we provide a more thorough characterization of this result. 

Although the $\Phi_*$-z trend that we find for the fiducial model is quite flat, due to the covariant nature of the 
Schechter parameters and the absolute magnitude range considered, drawing conclusions about the high-mass end from  
the Schechter parameters alone can be misleading. 
In any given scenario, the best way to provide a quantification of the evolution of the high-mass end of the LF is to 
define an absolute magnitude where a given number density is reached, and track the evolution of such 
magnitude as a function of redshift (see \citealt{Brown2007}, \citealt{Cool2008} and \citealt{Marchesini2014}).
As noted by \cite{Brown2007}, this method is more suitable for the high-mass end than studying 
the evolution of the luminosity density (see an example in \citealt{Wake2006}).
Here we define two absolute magnitudes, $M_{min}$ and $M_{max}$, so that $\log_{10}\Phi(M_{min}) = -5.5$ and $\log_{10}\Phi(M_{max}) = -4.5$, 
respectively. We also define the stretch parameter 
$\Delta M = M_{max} - M_{min}$, which is an alternative 
way of parametrizing the massive-end of the LF. The values of $\log_{10}\Phi(M_{min}) = -5.5$ and $\log_{10}\Phi(M_{max}) = -4.5$ are 
defined so that the corresponding $M_{min}$ and $M_{max}$ are brighter than the $50\%$ completeness limit
at any redshift, but are fainter, at any redshift slice, than the magnitude range where the data-model
deviations are large and small-number effects become important.

\begin{figure}
\begin{center}
\includegraphics[scale=0.39]{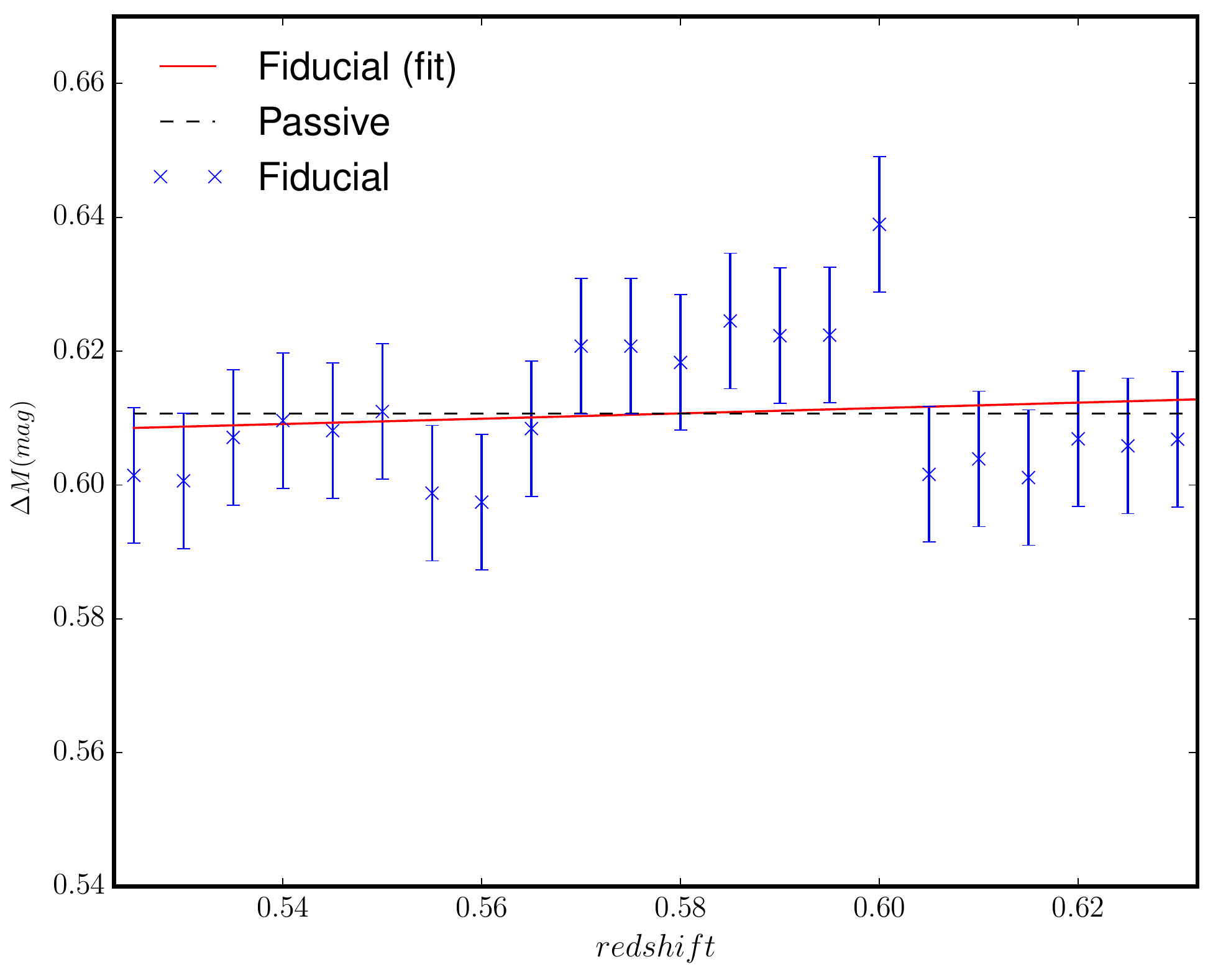}
\caption{Deviation from passive evolution. For the fiducial (crosses), the 
fit to the fiducial model (stars) and the best-fit passive model (which fades at a rate of 1.5 mag per unit redshift, see text), we show the 
stretch parameter as a function of magnitude. The stretching parameter $\Delta M$ is defined as 
the difference $M_{max} - M_{min}$, where $M_{min}$ and $M_{max}$ are the magnitudes
corresponding to the fixed number densities $\log_{10}\Phi(M_{min}) = -5.5$ and $\log_{10}\Phi(M_{max}) = -4.5$, 
respectively.}
\label{fig:stretching}
\end{center}
\end{figure}

In a passive-evolution (constant-$\Phi_*$) scenario, the stretch-parameter evolution is flat by definition\footnote{
Theoretically, one could also conceive a non-passive-evolution scenario where mergers and star formation conspire in a way that 
$\Phi_*$ remains constant, but this scenario seems very unlikely.}. 
In Figure~\ref{fig:stretching}, we show this parameter for the fiducial model, the fit to the fiducial model and the passive model
with  $\Delta ^{0.55}M_* /z \simeq 1.5$. Within a scatter of $\pm 0.01$ mag ($\lesssim 1 \%$ in flux), the significance of the non-passive trend is almost zero, 
as expected given the flat  $\Phi_*$-z trend. We have checked that above $z\sim0.63$, 
$M_{max}$ becomes very unreliable: the $M_{max}$ -z trend flattens completely, which is due to the fact 
that the effective magnitude range becomes too narrow.

Better characterizing the exact level of any merging and ongoing star-formation 
within this galaxy population will require a more accurate determination of the covariance matrix in follow-up work 
(which will also extend our analysis to a wider CMASS redshift range). An interesting result to follow up on
is the fact that our results are, in principle, consistent with a wide range of fading rates, some of which 
appear too high, as compared with previous literature (recall, however, that this is the first work that explores such a  
high-mass range at this redshift with significance). In fact, the observed, not-completeness-corrected RS LF (obtained applying a colour cut, not 
through any deconvolution method) shown in crosses in Figure~\ref{fig:lf_all} evolves at a rate of 
$\sim 2$ mag per unit redshift, at a number density of $\log_{10}\Phi(M_{min}) = -5$. Although our deconvolution method 
predicts plausible fading rates as low as $\sim1.1$ mag per unit redshift, this is an important aspect to clarify in future works.

\subsection{The fading rate of the RS LF as a tool for constraining the ages of the LRG population}
\label{sec:fading_rate}

Our results for the evolution of the RS LF are consistent with a passive-evolution scenario,
to the extent that the uncertainties in the modeling allow. 
Under such an assumption (or approximation), we can interpret
our constraints on $\Delta ^{0.55}M_* /z$ as constraints on the age of these populations.
Our approach is distinct from the methods of previous measurements, 
which are obtained from wide redshift ranges and usually a few 
($\lesssim3$) data points. For the first time (at least at these redshifts) we have enough statistics that we can divide 
our redshift range in extremely narrow redshift slices, and provide a fading rate (under the 
assumption of passive evolution), much in the sense of the local derivative of 
$^{0.55}M_*$ with respect to redshift. We note that this inference is only approximate,
given the range of possible $\Delta ^{0.55}M_* /z$ values that we find.

\begin{figure}
\begin{center}
\includegraphics[scale=0.45]{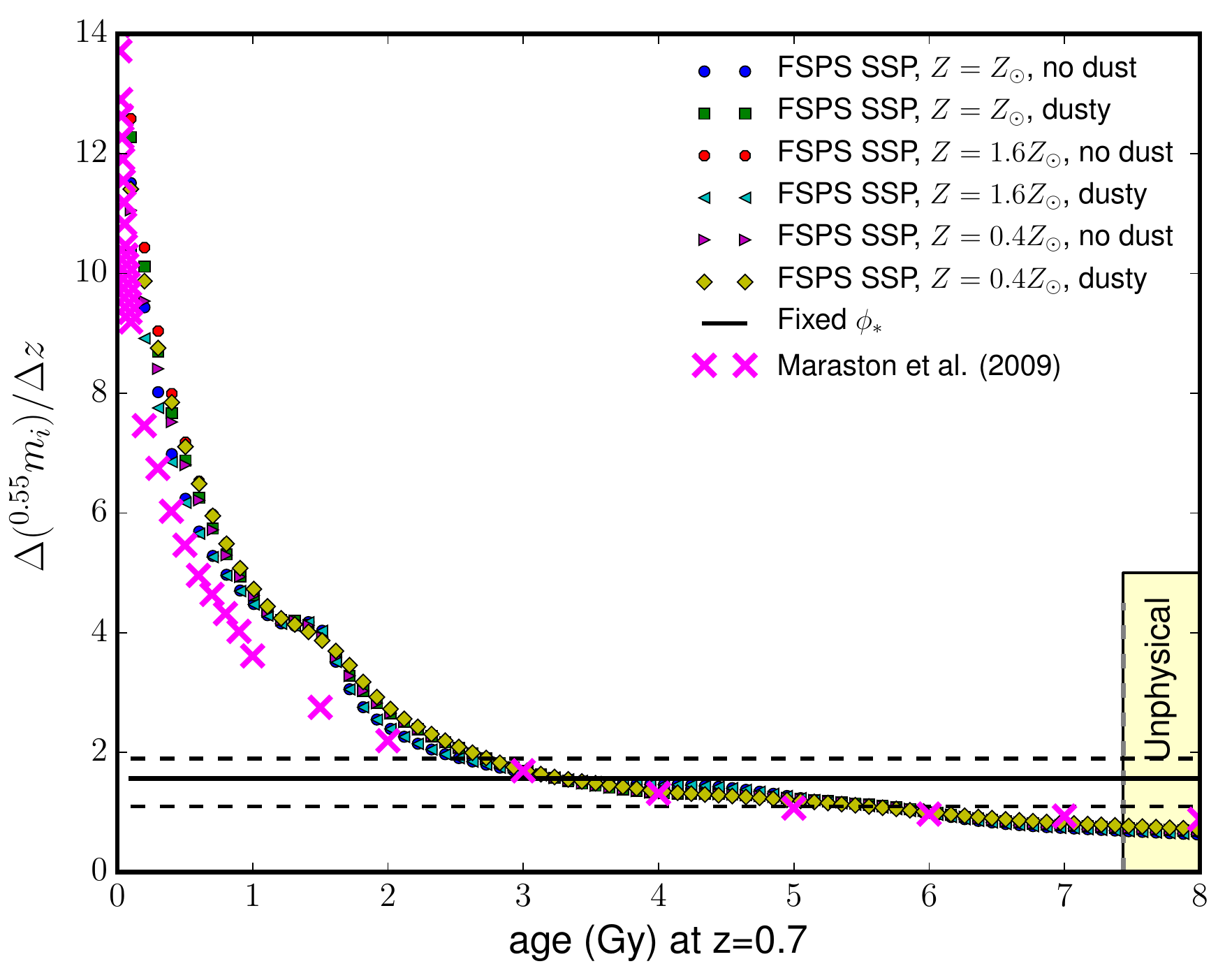}
\caption{The fading rate $\Delta ^{0.55}m_i / \Delta z$ for a set of 
SSPs as a function of the age at $z=0.70$. Symbols represent both FSPS models 
covering a wide range of stellar population parameters and the M09 passive 
LRG model. The horizontal lines represent a fading rate of $1.5\pm0.4$ mag per unit redshift.
These value corresponds to the fading rate within the redshift range 
$0.525 \leq z \leq 0.630$ for the best-fit passive-evolving fixed-$\Phi_*$ model.}
\label{fig:passive}
\end{center}
\end{figure} 

In Figure~\ref{fig:passive} we show the fading rate  
$\Delta ^{0.55}M_i / \Delta z$ for a set of single stellar population models (SSPs), as a function 
of the age at $z=0.70$. In the absence of mergers and star formation (passive), we can assume that an SSP
provides a good description of the star formation history (SFH) of the stellar population \footnote{
Note that it is common in literature to use the word ``passive" to describe the evolution of 
a galaxy population with no ongoing star formation, so the evolution is due to the fading 
of the stellar populations. In this context, a galaxy population undergoing mergers
can have a ``passively evolving stellar population'', provided that mergers do not result in 
new star formation.}.
In particular, we show both FSPS models and the \cite{Maraston2009} passive
LRG model (M09). For FSPS models, 
we show models for three different metallicities and with or without 
a moderate amount of dust. The fading rate is clearly a good tracer 
of the age of an SSP because it is insensitive to the model degeneracy. If we assume SSP $\tau$-models, the fading rate that we report
at $z\sim0.55$ would be consistent with a formation 
time for the LRGs of $\sim 3-5$ Gyr at $z=0.70$, which implies a formation redshift of $z\sim1.5-3$. 

Equal conclusions regarding the age of the LRG population are drawn when comparing our results with the 
M09 model, as Figure~\ref{fig:passive} shows. \cite{Maraston2009} were able to match 
the observed g, r, i colours of the LRG population across the redshift range $0.1<z<0.7$. In order to achieve this goal, 
the model requires $\sim$3 per cent of the stellar mass to be in old metal-poor stars. \cite{Maraston2009} also 
show that empirical stellar libraries such as the Pickles library \citep{Pickles1998} provide considerably better fits than theoretical libraries. 
 
Even though this is still a coarse measurement, the formation redshift (and age) that we measure using the fading rate of the RS LF is 
consistent with results from high-redshift quiescent red galaxies at $1.5\lesssim z \lesssim 2$ (using also SSPs). The 
spectroscopically-derived SSP-equivalent ages of these systems at their corresponding redshifts are $1-2$ Gyr 
(\citealt{Mendel2015}, \citealt{Whitaker2013}, \citealt{Onodera2014}, \citealt{Bell2004}), which 
correspond to a truncation redshift for the SSP model of $z_{trunc}=1.5-2.5$. This suggests that these high-redshift quiescent galaxies might be the progenitors
of the very massive LRG population at $z=0.55$ that we target in BOSS. As
Figure 4 from \cite{Mendel2015} shows, the ages that we measure at $z=0.55$ appear  
older than what has been previously measured at $z<1$ (see e.g. \citealt{Choi2014}), although 
tighter constraints are needed to confirm this scenario.

\subsection{The colour evolution of the high-mass end of the red sequence}
\label{sec:colour_evolution}

The intrinsic (g-i) colour of the RS at our reference magnitude can be well approximated
by the following linear relation:

\begin{equation}
(g-i)_{RS} = (3.217\pm0.074)~z + (0.996 \pm 0.042)
\label{eq:colour}
\end{equation}

within the approximate redshift range $0.5\lesssim z \lesssim 0.7$. Note that
these are observer's-frame, apparent colours. Intrinsic here means noise-free, i.e., 
relative to the intrinsic distributions. 

\begin{figure}
\begin{center}
\includegraphics[scale=0.37]{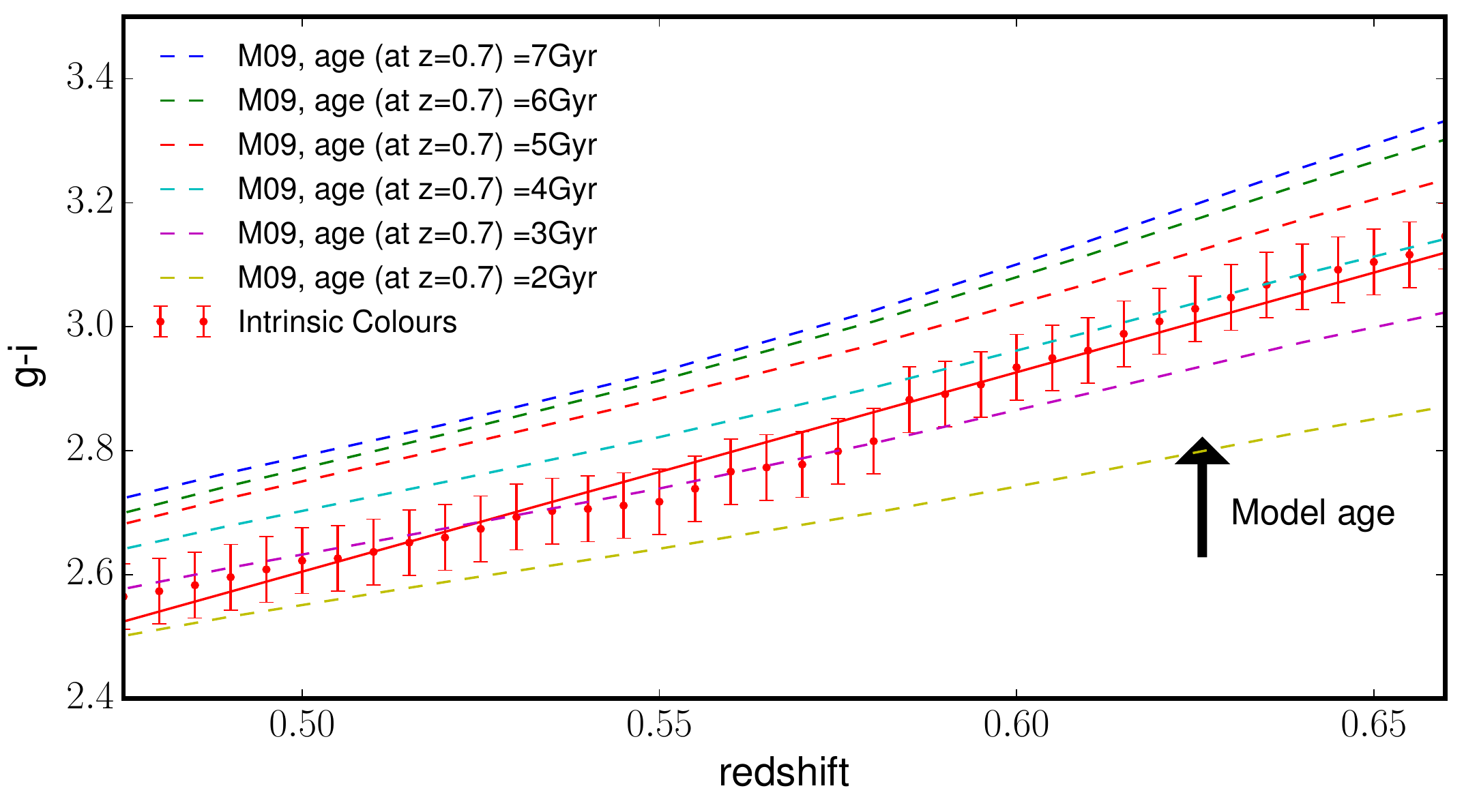}
\caption{The redshift evolution of the (g-i) centroid of the red sequence in intrinsic space, obtained using our fiducial
model, as compared with the M09 model. Due to the magnitude dependence of the centroid, we show the value at the 
reference magnitude $i=19$ ($Y_0+Z_0$). The arrow is added for the sake of clarity to indicate 
that dashed lines correspond to progressively older models as we move up in the plot. }
\label{fig:colour_evolution_m09}
\end{center}
\end{figure}

The evolution of the RS centroid is tightly connected to the evolution 
in the stellar populations of these systems. However, stellar population parameters are
very degenerate, in terms of their effect in a galaxy SED. Addressing the implications 
of our colour-evolution measurement requieres a detailed stellar-population analysis that is 
out of the scope of this paper. However, it is interesting to compare our results 
with the colours predicted by the M09 model, which is the result of a comprehensive stellar 
population fitting analysis specifically aimed at matching the 
{\it{observed}} colours of the LRG population as a function of redshift (see \citealt{Maraston2009}).
This comparison is shown in Figure~\ref{fig:colour_evolution_m09}, for the 
intrinsic (g-i) colour of the RS at our reference magnitude.

The intrinsic RS colours are consistent with models of ages between 3 and 4 Gyr at $z=0.70$, which 
is also consistent with the preliminary ages that we obtain from our fading rate measurement (3-5 Gyr).
Another way to look at this is the following: the M09 models that best fit the colours of the RS display 
a fading rate at $z\sim0.55$ of $1.3-1.7$ mag per unit redshift, which is consistent with our LF measurement.
Interestingly, the slope of the colour trend is in very good agreement with the predictions from M09.

In follow-up work, further constraints on the stellar population properties of RS galaxies 
can be placed by studying the evolution of average RS spectra. Our red+blue 
photometric population model can be extended 
to deconvolve average spectra of red and blue galaxies within the BOSS sample.

\section{Discussion}
\label{sec:discussion}

We have developed a method for deconvolving the 
intrinsic colour-colour-magnitude distributions from the effect of photometric errors, correlations between bands, and selection
in the BOSS CMASS sample. Our method, which is based on a forward-modeling approach, lays the foundation for future galaxy evolution studies 
using not only BOSS, but also future DE surveys such as eBOSS or DESI. The photometric deconvolution 
of spectroscopic observables is addressed in \cite{MonteroDorta2016}.

A key element in our deconvolution method is the error model derived from Stripe 82 multi-epoch data. 
By using this error model, the shape of the intrinsic distributions can be obtained directly from 
the data, without the need for making any strong assumptions based on stellar populations synthesis models. 
Our modeling is, therefore, not based on any arbitrary definition of ``red" and ``blue" galaxies,
but reflects the phenomenology of the colour-colour-magnitude diagrams (see a similar philosophy in \citealt{Taylor2015}).

Our results indicate that the high-mass RS at $z\sim0.55$ is consistent with a delta function
in the colour-colour plane, for a given magnitude and narrow redshift slice, {\it{within the resolution 
limit imposed by photometric uncertainties}}. We estimate that a width for the RS larger than $\sim 0.05$ mag
would be inconsistent with the observed scatter and our error model.
Below this approximate limit, adopting a non-zero value for the width of the RS has a negligible effect on our results.
Note that this description is, in principle, only valid at the bright end ($^{0.55}M_i \lesssim -22/-22.5$).
Importantly, in order to fit the number counts, our model requires a
colour-magnitude relation for the red sequence, with a slope of $\leq 0.05$ mag (colour) per unit mag. 

Our results  are qualitatively consistent 
with previous results for the RS. There is some consensus that the typical scatter in the colour-magnitude 
diagram of early type galaxies is significantly below $0.1$ mag at intermediate redshifts for the clusters red sequence: $\sim 0.05$ mag is 
the observed scatter measured by \citealt{Bower1992}, with at least $0.03$ mag due to observational errors. In the field, the 
intrinsic scatter is a factor 2 larger: $0.1-0.15$ mag (\citealt{Tanaka2005}, \citealt{Whitaker2010}, \citealt{Fritz2014}). 
The majority of previously reported colour-colour diagrams are presented for 
significantly wider redshift ranges and usually also larger magnitude bins (we use redshift intervals of $\Delta z = 0.01$). The combination 
of the colour-magnitude relation and the redshift effect can certainly account for the extra scatter measured 
by previous works. This is shown by comparing our colour-colour diagrams with those extracted from the PRIMUS survey at $z\sim0.5$. 

The high-mass BC, on the other hand, is defined within our modeling as the background component that is necessary to 
fit the observed distributions. This component is well described by an extended Gaussian function on 
the colour-colour plane, centered at $g-i\simeq1.7$. This Gaussian distribution intrinsically extends through the red side.
Note that by ``blue cloud" we do not necessarily mean ``young", ``late-type" or ``spiral", 
but a distribution of spectroscopically and photometrically heterogeneous objects that can be 
statistically separated from the narrow distribution that we call ``red sequence".
The fact that our model predicts that a fraction of BC objects would have intrinsically redder 
colours than the RS itself is in agreement with results from \cite{Taylor2015}.
We estimate that the fraction of intrinsically blue cloud objects on the red side of the colour-colour diagram is $\sim25-30\%$. 
This result is also quantitatively consistent with \cite{Taylor2015}, who find that a galaxy observed right on the locus of the R distribution still has 
a $20-25\%$ chance of having come from the bluer B population. The total fraction 
of intrinsically BC objects in the CMASS sample is $37\%$. 

From a morphological point of view, the fact that the CMASS sample is comprised by a significant fraction of non-RS galaxies is consistent with results 
from \cite{Masters2011}, who studied the morphology of 240 BOSS galaxies at $0.3<z<0.7$
using both HST imaging and COSMOS data. Quantitively, however, notable discrepancies exist. 
In particular, \cite{Masters2011} find that $29\%$ of the CMASS galaxies
in their 240-object sample did not present an early-type morphology; the majority are 
identified as late-type galaxies. Interestingly, only $\sim10\%$ of these objects 
lie on the red side of the $g-i=2.35$ demarcation, 
which would suggest a much more 
concentrated colour distribution for the BC, as compared to our model. Note that, however, uncertainties associated with  
mapping observed colours to morphologies are also well documented.

There is only limited resolution that can be attained 
using photometric data, so the inclusion of spectroscopic data into the modeling should help alleviate some of the 
above tensions. This aspect of the modeling, i.e., the connection between the RS and the 
BC and the role of the so-called {\it{green valley}} region in between, is key to understanding 
the way galaxies evolve (see examples of studies discussing these topics in \citealt{Bundy2010} and \citealt{Wolf2009}).

An important result of our analysis is the quantification of RS completeness 
in the CMASS sample,
as a function of colour and apparent magnitude. The sample is complete in apparent magnitude
below $i\simeq19.5$, where completeness starts to decline (the exact magnitude depends on 
the redshift slice considered). Overall completeness for the RS, integrated within the 
limits of the CMASS selection, remains above $\sim 0.8-0.85$ at $z\gtrsim0.53$. 
RS completeness is, by itself, a key result to the science that can be pursued with the CMASS sample. 

We have checked that uncertainties in our error model and systematics
have, qualitatively, little effect on the above results. In order to mitigate the effect of these
uncertainties in the computation of the RS LF, we have restricted our analysis 
to a redshift range around $z=0.55$, in particular from $z=0.52$ to $z=0.63$.
Even though our statistics are large, placing tight constraints on the evolution of the 
RS LF from such a narrow redshift range requires very high precision in the deconvolution 
procedure. For this reason we can only state that the evolution of the RS LF is {\it{consistent}} with that of a
passive-evolving model, given the uncertainties in the modeling. In particular, our results indicate that,
within the approximate absolute magnitude range $^{0.55}M_i\lesssim-22$, the RS LF is well described by 
a Schechter function of constant $\Phi_* = (7.248 \pm 0.204)\times10^{-4}$ Mpc$^{-3}$ mag$^{-1}$, passively fading 
at a rate of $1.1-1.9$ mag per unit redshift.

There is consensus in the literature that the effect of mergers on the evolution of massive RS galaxies is considerably less significant 
than that of their less-massive counterparts. However, different works differ in the extent that this 
evolution approximates that of a purely passively-evolving galaxy population. Note that 
the concept of pure passive evolution is only an approximation, as mergers have been observed 
involving massive RS galaxies (e.g. Perseus A). In practice, we use this concept 
to describe a situation where {\it{the occurrence of mergers and star formation within a given population is low enough that it 
does not affect the global evolution of the population, as measured from statistics such as the LF}}. 
If confirmed, our results would be in agreement with LF-evolution results 
from \cite{Wake2006} and \cite{Cool2008}. Using spectral analysis, \cite{Cool2008} also conclude that
the evolution in the average LRG spectrum also supports a purely passive evolution for the LRGs since $z\sim0.9$. 
This idea is reinforced by subsequent work from \cite{Maraston2013}. From a 
galaxy-clustering perspective, the evolution in the clustering properties of LRGs from the BOSS 
samples alone also appear to support the idea of a passively-evolving galaxy population (see \citealt{Guo2013, Guo2014}). 

Some authors, however, have reported results that indicate that the effect of mergers is noticeable 
within this population, when studying the LF or clustering. \cite{Brown2007} report small deviation from passive evolution 
for the red LF using the NDWF and Spitzer IRAC Shallow surveys.
Also, \cite{Tojeiro2012} conclude that the LRG population at $z<0.7$ evolves, although slowly, i.e., less than 2 per cent by merging, when 
the two samples are properly matched and weighted. From cross-correlation function measurements, \cite{Masjedi2006} find 
results that indicate that major dry merger may still play a role in the late evolution of LRGs. \cite{Lidman2013} also found 
a major merger rate of $0.38\pm0.14$ mergers per Gyr at $z\sim1$, for BCGs using a sample of 18 distant galaxy 
clusters with over 600 spectroscopically confirmed cluster members between them. Very recently, \cite{Bernardi2016} has reported that, while the 
evolution of the LF/SMF from BOSS to the SDSS is consistent 
with a passive evolution scenario, this appears not to be the case for the evolution of clustering. SDSS galaxies appear to be 
less strongly clustered than their BOSS counterparts, which seems to rule out the passive-evolution scenario.

Within the uncertainties of the study, we do not find conclusive evidence of non-passive evolution. 
From our detailed analysis of the BOSS data, we have identified several ways to improve our modeling, so 
a better characterization of the exact level of mergers and tar formation can be provided. 
First, we can attempt to reduce the uncertainty in our covariance matrix model, which will reduce 
the error bars in our Schechter parameters and allow us to explore a wider redshift range in BOSS. 
Second, we can combine our BOSS results with the SDSS, the BOSS LOZ sample or the eBOSS LRG sample. The techniques 
that we have presented in this paper lay the ideal framework to perform this study in a consistent way. This is a similar 
approach as the one followed by \cite{Bernardi2016}, but in the aforementioned paper the authors acknowledge that incompleteness in the BOSS sample is
not treated rigorously, which might be the cause of some of the tensions reported. Finally, we can explore 
the possibility of incorporating tighter {\i{a priori}} constraints on the BC from complementary 
datasets in a consistent way.

\section{Future applications: the intrinsic halo-galaxy connection}
\label{sec:future}

In \cite{MonteroDorta2016} we use our completeness/intrinsic distribution results
within a similar framework as the one presented here to measure the $L-\sigma$ relation at $z\sim0.55$.
These two works provide a detailed characterization of the high-mass RS population at 
$z\sim0.55$, that can be injected into a broader cosmological framework, in a consistent way.

In particular, our intrinsic characterization
can be used, in combination with N-body numerical simulations, to place constraints on 
the halo-galaxy connection. Importantly, {\it{this will provide a new perspective to BOSS clustering science,
which so far has entirely focused on observed distributions}}. In essence, our modeling 
can be combined with state-of-the-art publicly available N-body cosmological simulations,
to generate hyper-realistic mock catalogs for different cosmologies and where the 
halo-galaxy link is conveniently parametrized. The connection between galaxies and halos will be performed 
by applying the techniques of halo occupation distributions (HOD: e.g., \citealt{bw02, zeh05}) and halo 
abundance matching (HAM: e.g., \citealt{val04, tru11}). By applying our error model 
and the selection criteria to the resulting set of catalogs and fitting for 
the observed clustering properties and the observed LF, constraints on cosmological 
parameters and on the physically-motivated parameters of the {\it{intrinsic halo-galaxy connection}} model can be placed.

In addition to the cosmological applications, our redshift-dependent luminosity function can provide informative priors for the determination
of spectroscopic redshifts from low signal-to-noise survey data in future enhancements to the redshift 
measurement pipeline for BOSS and subsequent surveys \citep{Bolton2012}.  Also, we may use our red+blue 
photometric population model and the PDSO method to deconvolve average spectra of red and blue galaxies within the BOSS sample.
Finally, the forward-modeling techniques developed in this work can be applied to quantify the evolution of 
massive galaxies as a single population across redshift from $z \approx 1$ down to $z = 0$ by including other
galaxy samples from SDSS-I, -II, -III, and -IV. This can provide tighter constraints on the evolution of massive RS galaxies.

\section{Summary}
\label{sec:conclusions}

The DR10 CMASS sample of the BOSS survey contains $\sim600,000$ galaxies ($0.45\leq z \leq 0.70$)
covering, with unprecedented statistics, the very massive end of the RS population  
at $z\sim0.55$. Such a huge sample has great potential to shed light 
into the evolution of massive galaxies, but it also 
presents significant challenges. We have developed a forward-modeling method based on an 
unbinned maximum likelihood approach for deconvolving the intrinsic CMASS distributions
from photometric blurring and accounting for selection effects. Importantly, 
our approach can be applied to other future low-S/N dark energy surveys. 
The method allows us to: (1) determine the best-fit intrinsic (g-r) colour-(r-i) colour -i-band magnitude distribution for the red sequence at intermediate 
redshifts with unprecedented accuracy, including a correction for the contamination 
from blue cloud objects in the sample; (2) determine red sequence completeness in the sample as a function of 
colours and magnitudes; (3) compute the bright end of the RS LF.

The main conclusions of our analysis can be summarized as follows:

\begin{itemize}
	\item {\it{Our characterization of completeness in the CMASS sample lays the foundation for a variety of galaxy evolution and large-scale structure
	studies with BOSS}}	
	\item {\it{The high-mass red sequence at $z\sim0.55$ is extremely compact in the colour-colour plane. At fixed magnitude and 
	in a narrow redshift slice, the distribution is consistent with a single point, to the extent that photometric errors allow. In order to fit the observations
	the width of the RS cannot be larger than $\sim0.05$ mag. }}
	\item {\it{The high-mass blue cloud, defined as a spectroscopically and photometrically heterogenous population clearly distinguishable from the red sequence, can be modeled as
	a much more extended gaussian. Our intrinsic 	distributions reflect purely the phenomenology of the colour-colour plane.}}
	\item {\it{We have computed the RS LF in several redshift slices around $z=0.55$, i.e. within the redshift range $0.52<z<0.63$. Within this
	narrow range, the evolution of the RS LF seems consistent with passive evolution, within the uncertainties in the modeling.}}
	\item {\it{The fading rate measured from the RS LF implies a formation redshift of $z=1.5-3$ for the LRGs, under the SSP assumption. This 
	formation redshift is in agreement with that derived from spectroscopy studies of high-redshift quiescent galaxies ($1.5<z<2$), which suggests 
	that these galaxies might be the progenitors of massive LRGs at $z=0.55$.}}
	\item {\it{The characterization of the intrinsic properties of the high-mass red sequence at $z\sim0.55$ presented in this work and in \cite{MonteroDorta2016} has unique 
	potential to shed light into the intrinsic halo-galaxy connection for this galaxy population.}}
\end{itemize}

\section*{Acknowledgments}

This material is based upon work supported by the U.S. Department of Energy, Office of Science, 
Office of High Energy Physics, under Award Number DE-SC0010331.

The support and resources from the Center for High Performance Computing at the University of Utah are gratefully acknowledged.

AMD, FP, JC, CC, GF, SR thank the support from the Spanish MICINNs Consolider-Ingenio 2010 Programme under grant 
MultiDark CSD2009-00064, AYA2010-21231-C02-01, and MINECO Centro de Excelencia Severo Ochoa Programme under grant SEV-2012-0249.

Funding for SDSS-III has been provided by the Alfred P. Sloan Foundation, the Participating Institutions, the National Science 
Foundation, and the U.S. Department of Energy Office of Science. The SDSS-III Web site is http://www.sdss3.org/.

SDSS-III is managed by the Astrophysical Research Consortium for the Participating Institutions of the SDSS-III Collaboration 
including the University of Arizona, the Brazilian Participation Group, Brookhaven National Laboratory, University of Cambridge,
University of Florida, the French Participation Group, the German Participation Group, the Instituto de Astrofisica de Canarias, 
the Michigan State/Notre Dame/JINA Participation Group, Johns Hopkins University, Lawrence Berkeley National Laboratory, Max Planck
Institute for Astrophysics, New Mexico State University, New York University, Ohio State University, Pennsylvania State University, University of 
Portsmouth, Princeton University, the Spanish Participation Group, University of Tokyo, The University of Utah, Vanderbilt University, University
of Virginia, University of Washington, and Yale University.

\bibliography{./paper}

\appendix 

\section{Modeling the covariance matrix with Stripe 82}

We use Stripe 82 multi-epoch data to model the blurring effect produced by photometric errors and correlations between different {\it{ugriz}} bands.
This effect is described by the covariance matrix. Stripe 82 is the SDSS stripe along the celestial equator in the southern Galactic cap, 
covering the region defined by $-50^{\circ} < \alpha_{J2000} < 59^{\circ}$; $-1.25^{\circ} < \delta_{J2000} < 1.25^{\circ}$, i.e. 
a total of $\sim 270$ sq deg in the sky. What makes Stripe 82 particularly useful for this work is the fact that it was observed multiple 
times, as many as $\sim 80$ times before the final release of the Stripe 82 database (usually as part of 
the SDSS-II supernova survey, \citealt{Frieman2008}). The Stripe 82 database is comprised by a total of $303$ {\it{runs}} (continuous scans of the imaging telescope), 
plus two coadd runs. These runs correspond to different {\it{reruns}}, which is how the processing algorithm used is indicated. 
By imposing the latest Stripe 82 rerun, our Stripe 82 database is reduced 
to 120 runs (plus 30 extra runs which are not included in the official database but have some 
overlap with the Stripe 82 region). The total number of detections, of all types, in this data base exceeds 100 million.

\begin{figure}
\begin{center}
\includegraphics[scale=0.38]{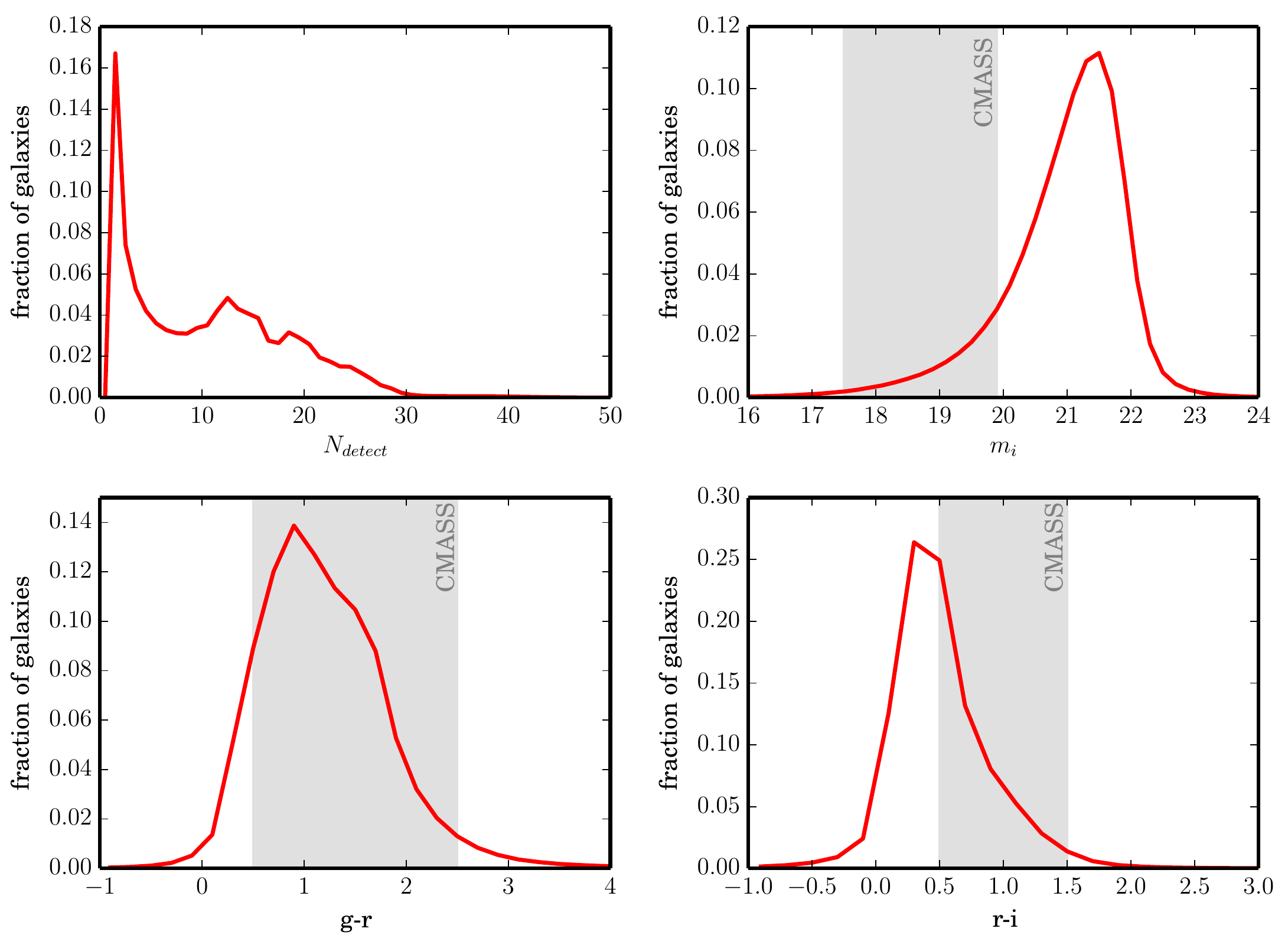}
\caption{The distribution of the number of detections ($N_{detect}$), the i-band magnitude, the g-r colour
and the r-i colour for t.pdfhe entire multi-epoch Stripe 82 summary catalog. In the last three panels, a shaded region highlights 
the approximate range covered by the CMASS sample.} 
\label{fig:stripe82prop}
\end{center}
\end{figure} 

In order to identify unique objects within the data base, we use the identification number THING$\_$ID, which is the same for all detections
of the same object across different runs. Each run in the database (and in the SDSS in general) is divided into a number of {\it{fields}}, 
this number varying for different runs. For each field, all the needed individual-detection photometric information is contained in the {\it{PhotoObj}}
files. Geometric and general photometric information for the corresponding field and run are stored in the upper-level  
{\it{PhotoField}} and {\it{PhotoRun}} files, respectively. A number of fields from different runs 
overlap in our Stripe 82 database, thus giving rise to multiple observations for the same object.

\begin{figure*}
\begin{center}
\includegraphics[scale=0.45]{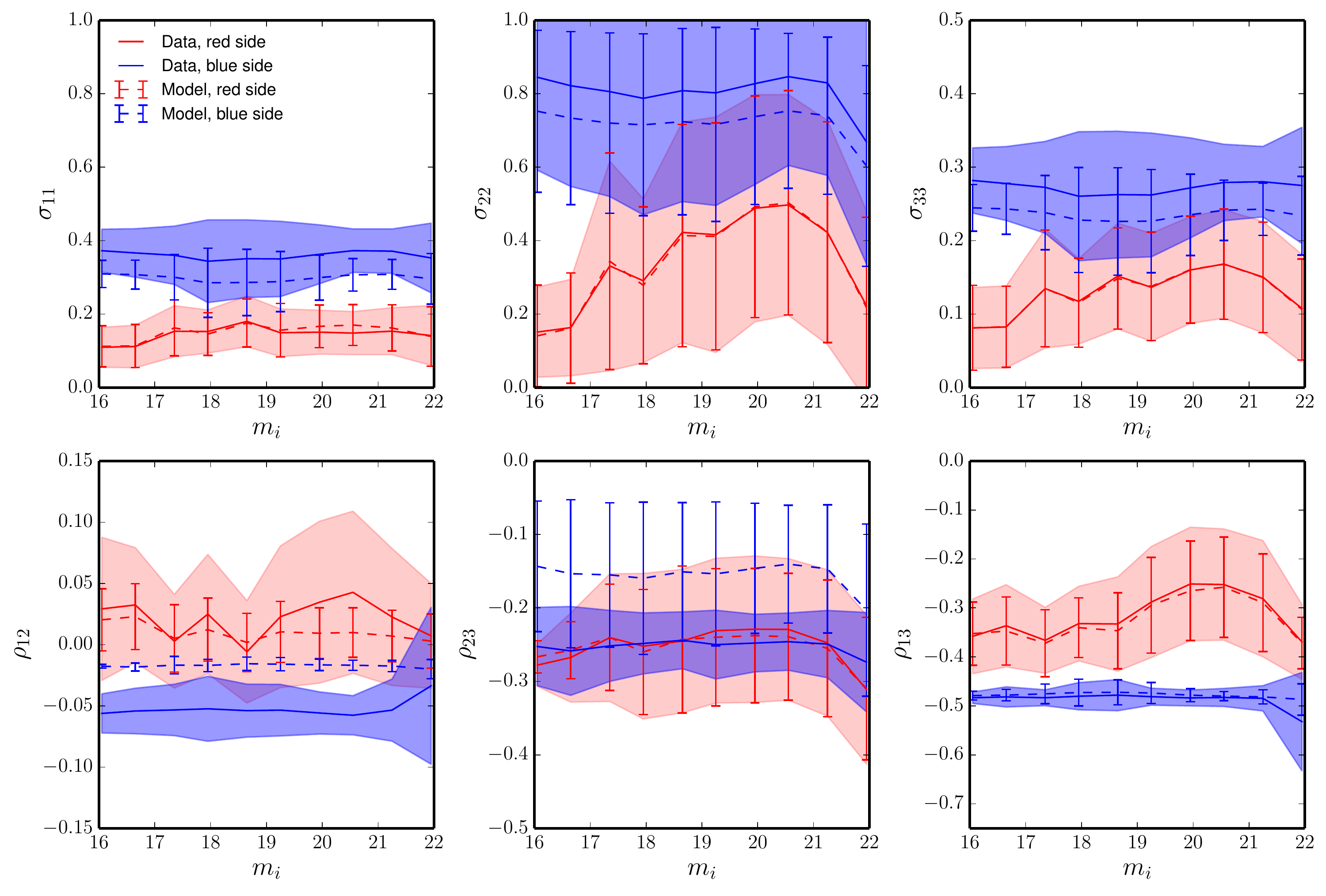}
\caption{Performance of the best-fit quadratic model for the dependence of the standard deviation terms $\sigma_{11}$, 
$\sigma_{22}$, $\sigma_{33}$ and the correlation terms $\rho_{12}$, $\rho_{13}$ and $\rho_{23}$ terms on colour and 
magnitude. The indices $1, 2, 3$ correspond to i, (g-r) and (r-i), respectively. In each panel, the red solid line and the red dashed line represent 
the average of the corresponding term in bins where $g-i \geq 2.35$ in the data and in the 
model, respectively. The shaded red region represents the 2-$\sigma$
scatter for the data, and the red error bars the corresponding 2-$\sigma$ scatter for
the model. In a similar format but in blue colours we show the average of each term for bins that 
satisfy $g-i < 2.35$. } 
\label{fig:covmatrix_fitting}
\end{center}
\end{figure*}

The THING$\_$ID identification number allows us to collect all detections for the same object, but the 
information about how each detection is resolved in the survey is stored in a bitmask called RESOLVE$\_$STATUS. 
Only objects with the SURVEY$\_$PRIMARY bit of the RESOLVE$\_$STATUS flag set
are used for the galaxy targeting algorithms. In order to collect a set of detections for each unique object, we must replace this condition.
Thus, we require all detections to have the RUN$\_$PRIMARY bit set. This criterion guarantees that detections are unique within each run. 

Finally, we requiere that all detections pass the SDSS photometric pre-selection criteria (\citealt{Dawson2013}, Section 2.2), including the photometric 
flag cuts and the star/galaxy separation. Note that the CALIB$\_$STATUS bitmask allows us to exclude all 
detections observed in non-photometric conditions, i.e., detections in fields that fail to satisfy the high photometric quality 
of the survey. Our final Stripe 82 multi-epoch summary catalog 
contains a total of $5,173,086$ objects that were ever selected (in any epoch) as
candidate galaxy targets. Each object has an average of $\sim10$ associated detections. For each object in the 
catalog, we compute the average i-band magnitude, g-r colour and r-i colour within the corresponding set of 
detections. We also calculate for each object the standard deviations of each variable
($\sigma_{11}$, $\sigma_{22}$, $\sigma_{33}$ for  i-band magnitude, g-r colour and r-i colour, respectively) 
and the correlations for the cross terms ($\rho_{12}$, $\rho_{13}$, $\rho_{23}$). These
elements are all that is needed to compute the covariance matrix for each individual object. 

In Figure \ref{fig:stripe82prop} we present some statistical properties of our Stripe 82 multi-epoch catalog. There is a large 
fraction of objects for which we only have one or two detections ($\sim 20 \%$). These objects are excluded 
from our analysis, as we set a number of detections threshold at $N_{detect} = 5$ for the computation of individual 
statistics. 

The covariance matrix shows some non-negligible dependence on magnitude and colours.
In order to model this dependence, we first split our Stripe 82 sample in 3D bins of magnitude, 
g-r colour and r-i colour, with a constant bin size of $\Delta X=0.1$, $\Delta Y=0.1$ and $\Delta Z=0.1$. In each 3D bin, 
we compute the median of magnitude, colours, standard deviation terms 
and correlation terms, as long as the bin contains at least 25 galaxies.  Finally, we assume a quadratic 
dependence on magnitude and colours for each of the three standard deviation terms and three correlation terms. 
As an example, the $\sigma_{11}$ term is modeled as follows:

\begin{eqnarray} \displaystyle
\sigma_{11}(X, Y, Z) = c_{XX}^{11}X^2+c_{YY}^{11}Y^2+c_{ZZ}^{11}Z^2+c_{XY}^{11}XY+ \nonumber \\
c_{YZ}^{11}YZ+ c_{XZ}^{11}XZ+c_X^{11}X+c_Y^{11}Y+c_Z^{11}Z+c_0
\label{eq:cov}
\end{eqnarray}

\noindent with 9 different coefficients.  Each of the six individual fits is performed only within the region defined by the following 
ranges: $16 < i < 22$, $0 < g-r <  3$ and $0 < r-i < 1.5$, which exceeds the CMASS selection region. 

In Table~\ref{table:covmatrix} we provide the average value and the standard deviation for the
six different statistics that we model. As expected, the photometric errors on 
$g-r$ are considerably larger than the photometric errors on the i-band magnitude and $r-i$. 
This effect is noticeable in Figure~\ref{fig:ccd}, as the scatter in the observed CMASS distributions
is larger along the g-r axis. Obviously, no correlation exists between 
the i-band magnitude and the g-r colour, and a negative correlation is present between the i-band magnitude 
and the r-i colour and between the g-r colour and the r-i colour. 

\begin{table}
\begin{center}
\small{
\begin{tabular}{cccc}
\hline
\hline
	       element &     $<element>$&    $\sigma(element)$  \\
\hline
	       $\sigma_{11}$&    +0.166 &     0.075   \\

	       $\sigma_{22}$&     +0.357 &     0.301 \\
       
	       $\sigma_{33}$&    +0.138 &    0.075  \\

	       $\rho_{12}$&   +0.006  &      0.048   \\

	       $\rho_{13}$&  $-0.266$ &    0.103    \\

	       $\rho_{23}$&  $-0.335$ &    0.113  \\          
         
\hline
\hline
\end{tabular}}
\end{center}
\caption{The mean value and standard deviation for the six different statistical quantities for which 
we model the colour and magnitude dependence (the standard deviation terms $\sigma_{11}$, 
$\sigma_{22}$, $\sigma_{33}$ and the correlation terms $\rho_{12}$, $\rho_{13}$, $\rho_{23}$).
The indices $1, 2, 3$ correspond to i, (g-r) and (r-i), respectively}
\label{table:covmatrix}
\end{table}

A visual inspection of the fits reveals that the quadratic model is capable of reproducing the general trends reasonably well.
In an attempt to illustrate this we present, in Figure~\ref{fig:covmatrix_fitting}, the dependence of each term on apparent 
magnitude, for two different sub-samples, defined by a colour demarcation at $g-i = 2.35$. The general conclusion is that the best-fit quadratic model reproduces
the average magnitude dependencies quite accurately at the red side of the colour-colour plane (at least 
up to $g-i \simeq 2$), but some discrepancies appear at the blue side. These 
discrepancies are consistent with the typical scatter in these relations.

Finally , we must account for possible overall inconsistencies in the photometric quality between the Stripe 82 and 
the SDSS footprint. To this purpose, we allow our covariance matrix to be multiplied 
by a scale factor that could in principle have a small dependence on magnitude but should be 
close to $1$. The value or functional form for this factor, that we call $\beta(i)$, will be empirically determined as part of the 
modeling of the intrinsic distributions. A visual inspection of the residuals, shows
that a constant value for $\beta$ has opposite effects towards each 
side of the CMASS magnitude range. In particular, it tends to make the 
distributions too broad at the bright end and too narrow at the faint end. In order to
avoid these effects, we adopt the following empirically-developed functional 
form for $\beta(i)$:

\begin{equation}
\beta(i) = \left[ \frac{1 + 10^{0.6(X - i_{0} + \Delta i)}}{1 + 10^{0.6(X - i_{0})}} \right]^{1/2}
\label{eq:esf}
\end{equation}

\noindent where X is the i-band apparent magnitude and $i_{0}=19.5$.

\label{lastpage}

\end{document}